\documentclass{aa} 
\usepackage{txfonts}

\usepackage{color}
\usepackage{graphicx}          
\usepackage{amsmath}           
\usepackage{amssymb}           
\usepackage{multirow}          
\usepackage{float}
\newcommand{\msol}{$M_\odot$}                   
\newcommand{\lsol}{$L_\odot$}                   

\newcommand{\adeg}{$^{\circ}$}                   
 
\newcommand{\um}{$\mu$m}                         

\newcommand{\mgsio}{Mg$_2$SiO$_4$}
\newcommand{\alo}{Al$_2$O$_3$}

\newcommand{\radmcd}{RADMC-3D}
\newcommand{\cobold}{CO$^{5}$BOLD}
\newcommand{\darwin}{DARWIN}
\newcommand{\optool}{\texttt{optool}}

\newcommand{\dustmodel}{\texttt{st28gm06n052}}
\newcommand{\gasmodel}{\texttt{st28gm06n056}}
\newcommand{\darwinmodel}{\texttt{M2n315u6}}

\begin{document}

\title{Asymmetries in asymptotic giant branch stars and their winds}
\subtitle{I. From 3D RHD models to synthetic observables}

\author{
    Joachim Wiegert \inst{1}
      \and
    Bernd Freytag \inst{1}
      \and
    Susanne Höfner \inst{1}
}

\institute{Theoretical Astrophysics, Division for Astronomy and Space Physics, Department of Physics and Astronomy, Uppsala University, Box 516, 751 20 Uppsala, Sweden\\ \email{joachim.wiegert@physics.uu.se, joachimwiegert.astro@gmail.com}
          }
\date{Received 22 March, 2024; accepted 17 July, 2024}

\abstract
{Asymptotic giant branch (AGB) stars are significant contributors to the metal enrichment of the interstellar medium. They have strong dust-driven winds that have their origin in regions close to the AGB star's surface, where dense dust clouds form.}
{In this methods paper, we adapted models from advanced radiation-hydrodynamical (RHD) simulations as input for radiative transfer software to create synthetic observables. A major goal is to describe an AGB star's non-sphericity and to simulate its effects on the surrounding dusty envelope.}
{We developed tools in Python to translate models of an AGB star and its dust-driven wind from 3D RHD simulations with \cobold\ into the format used for radiative transfer with \radmcd . We preserved the asymmetric shape of the AGB star by including the star as a `dust species' and by using temperature data computed in \cobold. The circumstellar silicate dust from the 3D RHD simulation is included using \mgsio\ opacity data in \radmcd\ with spatially dependent grain sizes. We compared images and spectral energy distributions (SEDs) created with \radmcd\ of a model snapshot with similar output made with a spherically symmetric stellar atmosphere from the 1D program \darwin\ and with a point source star in \radmcd .}
{Our \cobold\ model features substantial and clumpy dust formation just above 3.4\,au from the grid centre ($\sim 1\,R_\star$ above the star), and large-scale structures due to giant convection cells are visible on the stellar surface. With the properties of VLTI as a basis, we have created simple synthetic observables where the dust clouds close to the star and features on the stellar surface are resolved. The flux density and the contrast to the star are high enough that optical interferometers, such as the VLTI, should be able to detect these dust clouds. We find that it is important to include asymmetric stellar models since their irregular shapes, radiation fields, and their dusty envelopes even put their marks on spatially unresolved observables and affect the flux levels and shapes of the SEDs. The effects on flux levels can mostly be linked to the clumpiness of the circumstellar dust. In contrast, the angle-dependent illumination resulting from temperature variations on the stellar surface causes shifts in the wavelengths of the flux maximum, as shown by replacing the asymmetric star with a spherical one.}
{The methods presented here are an important step towards producing realistic synthetic observables and testing predictions of advanced 3D RHD models. With the model used here, we find that optical interferometers should be able to resolve thermal emission from dense clouds in the dust-formation zone close to an AGB star. Taking the angle-dependence of SEDs as a proxy for temporal variations in unresolved data, we conclude that not all variability observed in AGB stars should be interpreted as global changes in the sense of spherical models.}


\maketitle

\section{Introduction}

Dust plays several significant roles throughout the universe. Our focus here is on the substantial amount of dust that forms above asymptotic giant branch (AGB) stars and initiates strong dust-driven winds. These stars represent the late evolutionary stages of intermediate mass stars (0.8 to 8\,\msol ), and they produce elements that are injected by such winds into the interstellar medium (see, e.g., reviews by \citealt{hoefner2018, matthews2024}). 

Characteristic of AGB stars is variability on various time scales due to pulsations, immense convection cells, and thermal pulses. Convection brings up newly produced chemical elements from the layers around the core to the surface, where the gas is intermittently accelerated outwards by shock waves (caused by convection and pulsation), but the gas does not usually achieve escape velocity and may fall back to the stellar surface. However, if the temperature of the gas gets low enough (e.g. due its to altitude above the stellar surface), dust and dust-driven winds will form with mass-loss rates ranging from $10^{-8}$ to $10^{-4}$\,\msol\,yr$^{-1}$ as a result. Iron-free silicates (e.g. \mgsio ) are transparent enough to form relatively close to the surface of an AGB star (i.e. some 1 - 2 stellar radii above the stellar surface), and grains of sizes 0.1 to 1\,\um\ are efficient at scattering near-infrared (NIR) light, making them prime candidates for initiating dust-driven winds \citep{hoefner2008psvt}.

In recent years, significant progress has been made in understanding the mass loss of AGB stars due to dust-driven winds. However, many questions remain, including how variations in temperature and luminosity on the stellar surface affect wind formation, what the exact nature of the dust is (its composition, grain properties), how these aspects couple with wind properties (e.g. its terminal velocity, clumpiness), and how the large scales of the circumstellar environment are coupled with its smaller scales (i.e. how to connect observations that more easily probe large scales to theory that can describe small scales). In short, a complete theory is still lacking. Observations are just now starting to reach the level of detail that can confirm or disprove theories concerning the physics and dynamics of gas and dust just above the dust-formation altitude \citep{matthews2024}.

The aim of this study is to enable a comparison of the latest 3D simulations with observations of dust-forming regions of AGB stars and stellar surfaces. We developed methods to translate a radiation-hydrodynamical (RHD) \cobold\ model of an asymmetric AGB star and its surrounding gas and dust \citep{freytag2023} to be input for radiative transfer simulations with \radmcd\ \citep{dullemond2012}, from which we can create synthetic observables. 

The transition from spherical 1D to full 3D models has been shown to be important. Three-dimensional dynamical simulations allow for simultaneous inward and outward motions and the formation of dense dust clouds close to the star, due to transient local pockets of cooler gas just above the stellar surface \citep{freytag2023}. This effect seems to favour the formation of dust-driven outflows, compared to spherically symmetric models.

There are many observations of large-scale structures in the circumstellar environments of AGB stars. The Atacama Large Millimeter Array (ALMA) large program ATOMIUM observes a plethora of various circumstellar envelope morphologies, discs, spirals, and shocks \citep{decin2020, gottlieb2022}. They attribute many of the differences they find when compared to isotropic models (e.g. up to two times higher wind velocities than isotropic models) to possible (substellar) binary companions. 

However, binarity is far from the only reason behind disc-like envelope shapes, shells, clumps, or other asymmetric wind morphology. For example, \citet{homan2021} studied the special case of R\,Hya in the ATOMIUM data. They found two bubbles in opposite directions around the star that they estimated to be connected to the previous thermal pulses of this star. Recently, \citet{velillaprieto2023} presented ALMA data of CW\,Leo, and they were able to reach an angular resolution that corresponds to the star's radius and thus resolve gas clumps within $\sim 10\,R_\star$ from the star, which seem to originate from convection in the star. In another example, \citet{coenegrachts2023} observed and modelled NaCl around the AGB star, IK\,Tau, and used the 3D radiative transfer programme MAGRITTE \citep{deceuster2020a, deceuster2020b} to compare the models with the observations. They also found non-spherically symmetric shapes such as spirals and several clumps within 100\,au of the star but with unclear origins. 

Recent observations have even been able to probe the regions where dust and winds are formed around AGB stars. \citet{montarges2023} did follow-up observations of the ATOMIUM sources with the Very Large Telescope (VLT) and modelled the dust (including polarisation) around these stars with \radmcd . They found clumpy dust structures in the inner parts of the envelopes, where dust formation occurs. Also \citet{khouri2020} observed asymmetric envelopes of AGB stars in polarised light with the VLT Interferometer (VLTI) for a few objects. The dust contents and the extent of wind acceleration around these stars fit well with theoretical predictions according to their findings. These environments were found to be dominated by clumps and arcs rather than shells \citep[see also][]{ohnaka2017}. Most recently, with a combination of new and archival data from VLTI at around 3 and 10\,\um\ of V\,Hya, \citet{planquart2024} have clearly observed dust emission from asymmetric clumps close to and associated with V\,Hya.

On even smaller scales, \citet{khouri2016} were able to resolve large-scale stellar surface features of R\,Dor with VLT, \citet{paladini2018} have resolved stellar surface features with VLTI of $\pi^1$\,Gruis, and more recently, \citet{rosalesguzman2024} have used VLTI to resolve convection cells on the surface of R\,Car. The example with R\,Dor was performed with a significantly lower angular resolution but with variations over a time span of 48 days, and \citet{rosalesguzman2024} used high-resolution VLTI data of R\,Car from two different epochs, 2014 and 2020.

An M-type AGB star may form magnesium-silicate dust (\mgsio , MgSiO$_3$) and alumina (\alo ) relatively close to the star ($\gtrsim$ 1 - 2\,$R_\star$ above the stellar surface; see, e.g., \citealt{gobrecht2016, hoefner2016, millar2016}). Enrichment of magnesium-silicates with iron can happen further out from the star \citep{hoefner2022}. 

Silicates have two characteristic infrared features at the wavelengths 9.8 and 18\,\um\ ($\sim$\,10 and 20\,\um , hereafter). The strength of these features depends on both the amount and temperature of the dust. The latter is set by radiative heating, depending on dust grain composition and optical properties. Three-dimensional modelling allows for significant local temperature variations that can result in more visible silicate features. Furthermore, using the same total dust mass of a circumstellar environment while changing the morphology of the envelope from spherical to disc-like shapes and varying the observation angle will significantly change how the overall dust spectral energy distribution (SED) appears. For example, along an edge-on line of sight (LOS), the observer may just observe the outer cooler dust and detect no silicate features \citep{wiegert2020}. As such, it is common to observe AGB stars for which it is difficult to fit SEDs based on 1D models \citep[e.g. recent studies by][]{groenewegen2022, olofsson2022}. 

Here, we focus on dust within some 10\,au from the stellar surface of an M-type AGB star in a 3D model produced with \cobold . We translate this model to be an input model for radiative transfer in \radmcd\ -- with which we produce synthetic observables. When working with synthetic observables based on 3D star-and-wind-in-a-box models, a question naturally arises regarding how much of what is seen in images and unresolved data (e.g. SEDs) is due to the morphology of the circumstellar environment (clumpy clouds), as opposed to non-spherical illumination caused by the patchy surface of the star. Therefore, we also present comparisons where we exchange the asymmetric star of the 3D model with a spherically symmetric AGB star based on a model from the 1D code \darwin.

This paper is structured as follows: In Sect.\,\ref{sec:codedescript}, we summarise the software and specific models that are used in this study (\cobold , \darwin , and \radmcd ). In Sect.\,\ref{sec:methods}, we describe our methods, namely, how we translate data from RHD simulations to be used in radiative transfer. In Sect.\,\ref{sec:results}, we present our simulation results, and we discuss differences caused by exchanging the star with a spherically symmetric star. In Sect.\,\ref{sec:observables}, we create proof-of-concept observables from the \cobold\ model, and Sect.\,\ref{sed:conclusions} contains our conclusions.

\section{Code and model descriptions}
\label{sec:codedescript}

\subsection{\cobold}

The code \cobold\ (``COnservative COde for the COmputation of COmpressible COnvection in a BOx of L Dimensions with $l = 2$ ,3''; see \citealt{freytag2012, freytag2013, freytag2017aap, freytag2017memsai} for detailed descriptions) is a fully 3D RHD code that can be used to simulate dynamical processes in stars and their surroundings. The resulting models include pressure waves, acoustic modes, and convection of the interior as well as shocks in the atmosphere of the star. The code computes radiative energy transport through the optical thick interior of the star and the optical thin atmosphere. Most recently, \citet{freytag2023} presented `star-and-wind-in-a-box' models with an extended box with a side length of $\sim 30$\,au that also includes dust formation and dust-driven winds. 

In their current version, the simulations describe the formation of magnesium-silicate grains (\mgsio ), using a gas-kinetic model of grain growth \citep[see][Sect.\,2.3 for details]{freytag2023}. The wind-driving radiation pressure acting on these silicate grains is described in their Sect.\,2.4. In short, \cobold\ computes the dust opacity that causes radiative acceleration on the fly, based on current grain sizes, during simulations. The models use an analytical approximation of this opacity, as introduced by \citet{hoefner2008psvt}, \citep[Eq.\,4 of][]{freytag2023}, guided by Mie theory. In the \cobold\ models, each spatial grid cell contains one specific grain size.

\subsection{\darwin}

To explore effects of illumination vs. intrinsic 3D structures of the dust clouds (Sect.\,\ref{sec:spherecompare}), we use the following test setup: The asymmetric star resulting from the \cobold\ simulation is replaced with an extended, spherically symmetric stellar source, while keeping the 3D dust distribution unchanged. The radial structures of the physical quantities that define the stellar atmosphere are taken from a model produced with the 1D RHD code \darwin\ (Dynamic Atmosphere and Radiation-driven Wind models based on Implicit Numerics, see \citealt{hoefner2016}, and references therein). 

In a \darwin\ model, the variable radial structure of the stellar atmosphere and wind results from solving the coupled system of gas dynamics, radiative processes, and non-equilibrium dust formation. In contrast to the 3D \cobold\ simulations that include the convective, pulsating stellar interior, \darwin\ models cover a region with an inner boundary below the stellar photosphere but above the driving zone of the pulsations. The effects of stellar pulsation on the atmosphere are simulated by temporal variations in gas velocity and luminosity at the inner boundary of the model. They are introduced gradually by increasing the amplitudes up to the full values, starting from a hydrostatic, dust-free, atmospheric structure corresponding to the fundamental parameters of the star. Following this initial phase, the simulations are run for several hundred pulsation periods to avoid transient effects. The resulting time series of snapshots (describing the dependence of velocities, densities, temperatures and dust properties on distance from the stellar centre) can be used as a basis for computing synthetic spectra and other observables \citep[see, e.g.,][]{hoefner2022}.

\subsection{\radmcd}

\radmcd\ is a Monte Carlo-based radiative transfer program written by \citet{dullemond2012} that recently reached v.\,2.0. It works in full 3D geometry and can handle arbitrary dust spatial distributions, with any (spatially overlapping) dust species, illuminated by any number of stellar light sources, and with the possibility of adding molecular lines. It is openly available online\footnote{\url{https://www.ita.uni-heidelberg.de/~dullemond/software/radmc-3d/}} and the source code is continuously updated on \texttt{GitHub}\footnote{\url{https://github.com/dullemond/radmc3d-2.0}}.

With \radmcd , one can compute scattering and absorption by dust, and dust radiative equilibrium temperatures. Regarding scattering there are several modes included; isotropic, anisotropic scattering approximation by \citet{henyey1941} with scattering phase functions, and fully polarised using scattering matrices (Muller matrix) in the dust species data. \radmcd\ computes grain temperatures for dust with non-grey opacities that is heated by either black bodies or stellar sources with any chosen input SEDs. The temperature is based on the assumption of radiative equilibrium.

It should be noted that \radmcd\ only includes the functionality to use point sources, extended spherical stars, or distributions of stars. However, all these options have various advantages and drawbacks that are not compatible with our study. Therefore, we had to develop a method to emulate a non-spherical star by other means, as described in detail in Sect.\,\ref{sec:stardescript}.

\subsection{Model descriptions}
\label{sec:modeldescript}

\begin{table*}
    \caption{Stellar properties of models included in this study.}
    \label{tab:includedmodels}
    \begin{center}
    \begin{tabular}{ccccccc}
\hline\hline
\noalign{\smallskip}
    Designation & Pulsation period (d) & $M_\star$ (\msol ) & $R_\star$ (au ) & $L_\star$ (\lsol ) & $T_{\rm eff}$ (K) \\
\noalign{\smallskip}
\hline
\noalign{\smallskip}
  \dustmodel$^a$   &  545   &  1.0  & 1.65  & 7030 & 2806\,K \\
  \gasmodel$^a$    &  390   &  1.0  & 1.36  & 5085 & 2848\,K \\
  \darwinmodel$^b$ &  490   &  1.5  & 1.67  & 6498 & 2730\,K \\
  Point source$^c$ &  N/A   &  N/A  & 1.65  & 6975 & 2800\,K \\
\noalign{\smallskip}
\hline\hline
\noalign{\smallskip}
\multicolumn{7}{c}{Additional model parameters} \\
\noalign{\smallskip}
\hline
\noalign{\smallskip}
               & Snapshot time (yr)   & Box size (au) & Number of cells & Includes dust & & \\
\noalign{\smallskip}
\hline
\noalign{\smallskip}
  \dustmodel   &  29.95               & $29.8^3$      &  $679^3$ & Yes    & & \\
  \gasmodel    &  22.02, 22.18, 22.34 &  $7.3^3$      &  $317^3$ & No     & & \\
  \darwinmodel &  361.64              &  46.8$^d$     &  98$^d$  & No$^e$ & & \\
\noalign{\smallskip}
\hline
    \end{tabular}
    \end{center}
    \begin{list}{}{}
    \item[Notes.] 
    $^a$ Stellar mass, radius, bolometric luminosity, effective temperatures are all simulation time averages as computed with \cobold .
    $^b$ This model is from 1D \darwin\ simulations and the snapshot was chosen to have luminosity and radius close to the \cobold\ model star. The effective temperature is given by Stefan-Boltzmann's law.
    $^c$ Point source refers to input stellar parameters for the \radmcd\ test star. Radius and temperature are input values, the luminosity comes from Stefan-Boltzmann's law.
    $^d$ The size and number of cells of the \darwin\ model refer to a 1D radial structure (centre to outer edge). The 1D model also uses a self-adaptive grid, concentrating points in critical regions.  
    $^e$ The original \darwin\ model contains dust that is not included in this study.
    \end{list}
\end{table*}

In Table\,\ref{tab:includedmodels} we summarise the stellar parameters and other simulation parameters for the models we used in this study. We note that these are averages and the properties we extracted from the radiative transfer runs will vary with snapshot time and observed direction. The snapshot time in the table refers to the simulation time in \cobold , or \darwin , of the snapshot we used here.

\begin{figure*}
    \centering
    \includegraphics[width=188mm]{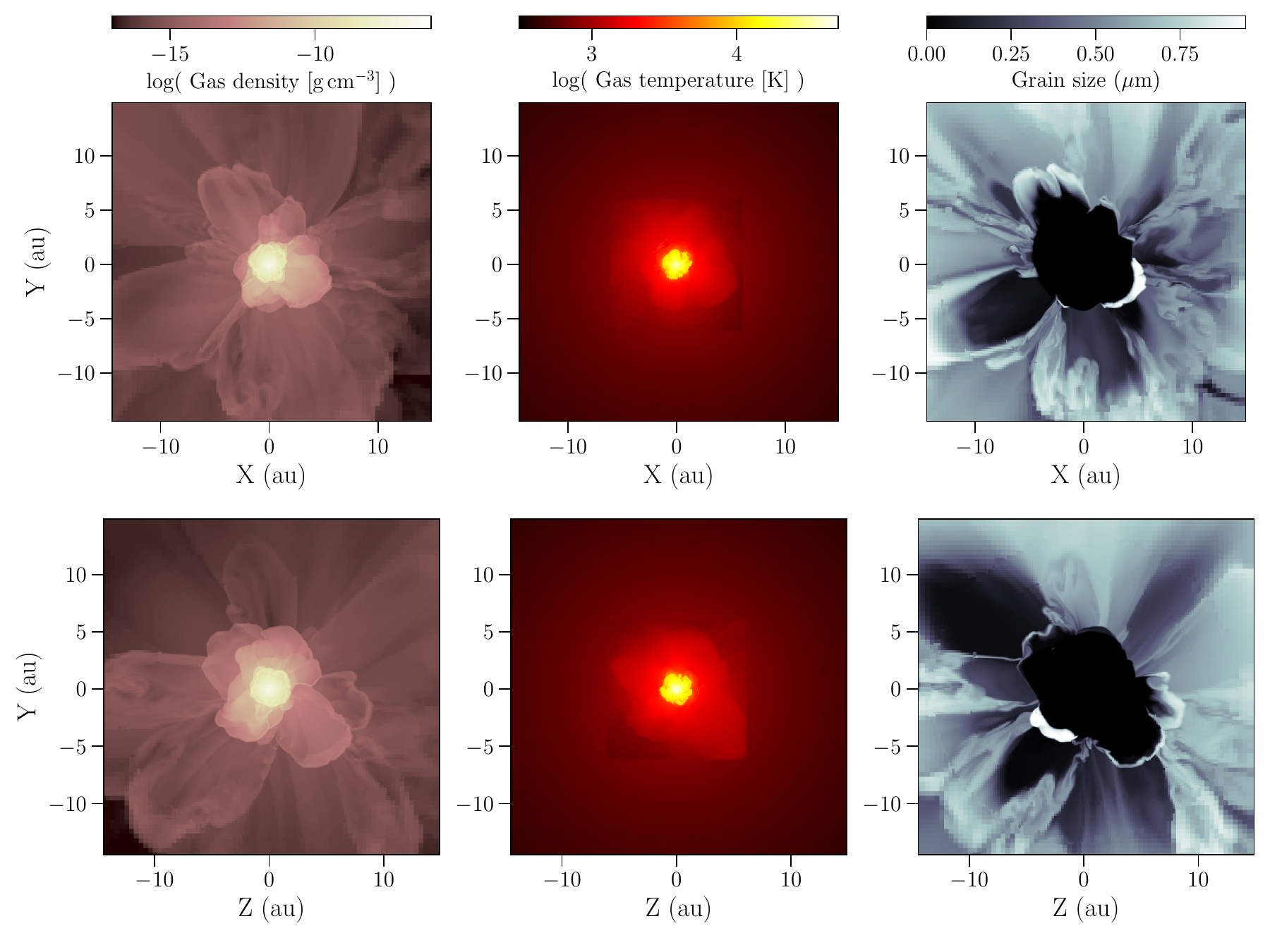}
    \caption{Images of a cut through the centre of the grid of a snapshot of model \dustmodel\ mapped to the \radmcd\ grid described in Sect.\,\ref{sec:griddescript}. The first row shows a 2D slice of the X-Y plane, and the second row shows the Y-Z plane. The left panels show gas densities, the middle panels show gas temperatures, and the right panels show grain sizes.}
    \label{fig:c5dslice}
\end{figure*}

The \dustmodel\ model is our main test object from \cobold\ and it was previously published by \citet{freytag2023}. Our choice of snapshot is at 29.95\,yrs of simulated time and is close in time to their 29.962\,yrs snapshot (marked as yellow in their Fig.\,4). It contains the simulation of convection through an M-type AGB star with a current mass of 1\,\msol , outflows of gas, dust condensation (\mgsio ), and wind acceleration within a box with the side of 29.8\,au. In Fig.\,\ref{fig:c5dslice} we show 2D slices through the centre of the box of model gas densities, temperatures and grain sizes. The average radius of the star is equal to $\sim\,1.65$\,au. The lowest altitude for dust formation at the included time snapshot corresponds to 3.4\,au, or 2.1\,$R_\star$, from the grid centre. Our snapshot is approximately at phase 0.9 of one luminosity-radius pulsation period (peak to peak).

We also used three snapshots of an alternative dust-free but otherwise similar model, designated as \gasmodel , when first developing and testing the translation codes from \cobold\ to \radmcd . It contains only gas data of an M-type AGB star with similar properties as the one in \dustmodel , but in a smaller computation box (side length of 7.34\,au).

The model designated as \darwinmodel\ is a 1D model from \darwin\ of an M-type AGB star as described by \citet{hoefner2022}. We used the gas density, temperature, and average Rosseland opacity from it to create a spherically symmetric stellar model in \radmcd . The data were extracted from one time snapshot of one pulsation cycle of this star with a period of 490 days. This snapshot corresponds to $\sim 0.3$ of one pulsation cycle (peak to peak) and is where the star has the most similar radius to the star from \dustmodel . With this we could study differences caused by the inclusion or exclusion of stellar surface features. 

Finally, the \textit{Point source} model is a standard black-body point-source star in \radmcd . This was used to compare with the extreme case of not including any spatial extent of the star at all. In \radmcd , the input values for a point source star are its physical position, effective temperature, and radius, and these in turn define its luminosity.

\section{Adapting radiation-hydrodynamical model data for \radmcd }
\label{sec:methods}

Here we describe how we adapt \cobold\ models as input data for \radmcd . Spatial grids are described in Sect.\,\ref{sec:griddescript}, gas data translation is described in Sect.\,\ref{sec:stardescript}, dust data translation in Sect.\,\ref{sec:dustdescript}, and dust grain properties and opacities are described in Sect.\,\ref{sec:opacdescript}.

\subsection{Grids: Spatial and wavelength}
\label{sec:griddescript}

In \cobold , the spatial grid consists of cubic and rectangular-shaped grid cells. The smallest grid cells are cubic and spatially limited to within an inner computation box around the centre of the grid and the star. In \dustmodel , this space has a side length of $\sim 12.22$\,au while the cells within this space are $\sim 0.026$\,au on their sides. Grid cells outside the inner box are increasing in size with distance to the grid centre and have mostly rectangular shapes since their sizes are limited by the sizes of the inner box' grid cells.

\begin{figure}
    \centering
    \includegraphics[width=70mm, angle=90]{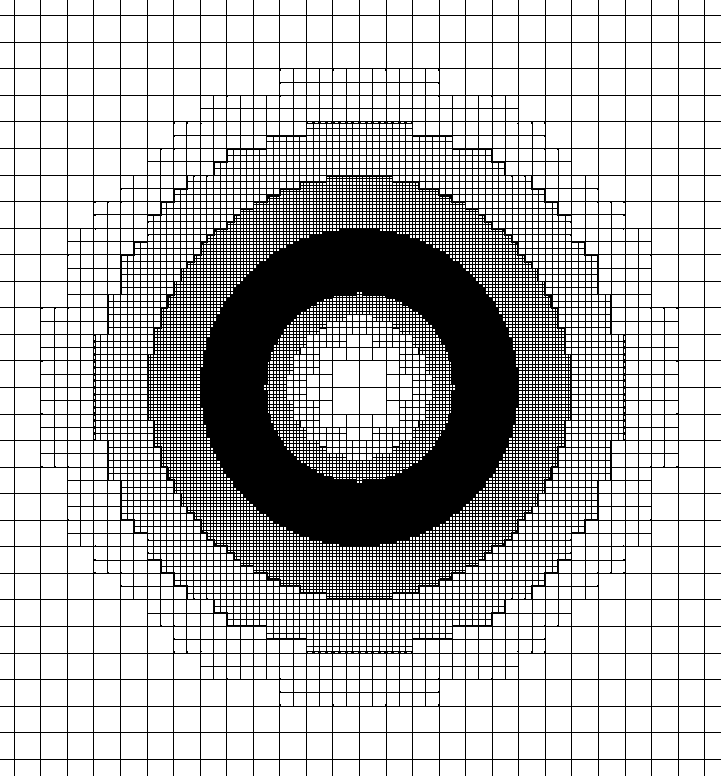}
    \caption{Schematic overview of the octree refined \radmcd\ grid we used. We lowered the resolution within the star and further out and kept the grid resolution high around the stellar surface and its surroundings.}
    \label{fig:radmcdgrid}
\end{figure}

For \radmcd , we used a cubic octree grid with four levels of refinement at various radial distances from the centre of the grid. A schematic example is shown in Fig.\,\ref{fig:radmcdgrid}. The smallest grid cells were set to be equal to those of \cobold 's smallest cells while, subsequent cell sides doubled in size for each refinement level. This gave five different cell sizes up to the base cells (largest cells) that have side lengths of 0.41\,au (0.25\,$R_\star$), and the grid was divided into $72^3$ base cells in the \dustmodel\ model. In total, this model has 19590712 cells in \radmcd . Because \cobold\ and \radmcd\ both use cubic grids it was relatively simple to translate data from one grid to the other even though their grid refinement styles differ.

In contrast to \cobold\ simulations, where the inner parts of the star and its convection zones are of utmost importance, these parts are not visible in the radiative transfer simulations since the stellar interior is optically thick. Therefore, we improved the performance by reducing the grid resolution inside the star. Thus we have radial ranges for each grid refinement from below the stellar surface, up to some distance above the star, while not reducing the grid resolution in extended regions of the stellar photosphere, and to retain a higher level of detail where dust is formed.

The decrease of the grid resolution in \radmcd , outside the star, is limited by the grid resolution of the \cobold\ model. Here, we set the \radmcd\ radial grid refinement limits so that its grid resolution was never higher than that in the \cobold\ grid.

Finally we also defined a wavelength grid for \radmcd . This was set to have the range of 0.1 to 500\,\um\ with 1000 logarithmic steps. This is a wide wavelength range that includes ultra-violet, wavelengths corresponding to the highest flux density of the star in visual (V) and NIR, primary wavelength regions for emission of circumstellar dust in NIR and mid-infrared (MIR), and finally the lower limits of the sub-millimetre region. Wavelength ranges that correspond to certain bands where current telescopes can observe circumstellar envelopes are of particular interest, for example, the European Southern Observatory (ESO) VLTI,\footnote{\url{https://www.eso.org/sci/facilities/paranal/telescopes/vlti.html}} which has instruments that cover wavelengths from $\sim\,1.6$\,\um\ to 50\,\um .

\subsection{Stellar and gas properties}
\label{sec:stardescript}

\subsubsection{Asymmetric star from \cobold }

The stellar structures from \cobold\ were saved in 3D arrays as densities of gas (in g\,cm$^{-3}$), Rosseland opacities (denoted $\kappa_{\rm Ross}$ in units of cm$^2$\,g$^{-1}$), and temperatures (in K). The $\kappa_{\rm Ross}$ data are constant with wavelength so all gas-filled grid cells emit as black bodies. In practice, this meant that we treated the star as a large number of black-body emitting cells.

To be able to simulate a non-spherical star in \radmcd , we treated the \cobold\ gas density, temperature, and opacity similar to a dust species. The temperature was immediately usable by \radmcd\ while the gas density and opacity required further processing, accounting for the spatial variations of the gas opacity. In \radmcd , for dust, one normally does this by separating the dust into separate species where each species has a file containing wavelength dependent data on absorption and scattering (and polarisation if required). However, since we used $\kappa_{\rm Ross}$ for the gas we had no dependence on wavelength, and could combine the gas density and Rosseland opacity to $\chi (\vec r) = \kappa_{\rm Ross}(\vec r) \times \rho (\vec r)$. This means that the gas' extinction input data in \radmcd\ was set to $\kappa_{\rm abs} = 1$ and zero scattering for all wavelengths, while the gas density input data in \radmcd\ was replaced by $\chi (\vec r)$, an opacity per grid cell.

A known problem that can occur when using Monte Carlo radiative transfer is the leakage of photons from hot optically thick regions to their optically thin surroundings. In our case, this issue could appear both as significantly brighter and hotter stellar SEDs than expected, and also as strong numerical noise when combining gas and dust data. For example, the stellar SEDs as simulated with \radmcd\ would be a few 1000\,K above 2800\,K, and some $10^4$\,\lsol\ instead of the expected $\sim 7000$\,\lsol . The numerical noise appeared as strong flux densities at random and narrow wavelength ranges. These effects were affected by choice of spatial grid resolution and seed number for random walk computations. 

This problem was also described by \citet{pinte2009} in their tests of Monte-Carlo codes for circumstellar discs in the NIR and MIR wavelength regimes (e.g. \texttt{TORUS, MCFOST, MCmax}). They found that scattered photon packages from the inner, dense, and hot regions of their discs dominated the simulated emission, and that the output spectra were sensitive to the choice of the grid resolution. Only a few packages may escape optically thick regions, but if they do, and the outer parts of the region is of much lower temperature than the inner parts, they can dominate the simulated images (private communication with C.\,P.\,Dullemond). This results in increased brightness and, in some cases, increased emission temperature in the image.

There are several settings in \radmcd\ to handle the behaviour of the random walk in a dense medium, for example, how easily a photon package is absorbed, or how photon packages are handled when stuck inside an optically thick region (see \texttt{tau\_abs} and ``Modified Random Walk'' in the \radmcd\ manual). Changing these settings did not result in any noticeable changes in our images and SEDs. Since the radiative transfer through the dust envelope should be as consistent as possible, and the non-spherically symmetric surface of the star is an essential part of the 3D model, we could not modify the random walk itself anyway. In that case, we may lose effects caused by optically thick dust and local variations in the density and temperature of the star. 

One possible solution to this problem is to increase the number of photon packages used to calculate scattering to an impractically high value. Another would be to disregard scattering altogether, which is not very useful either. Instead, we found that smoothing local peaks in temperature, density, and opacity within the star decreased this problem. Furthermore, we removed numerical noise by combining simulated SEDs with different random seed numbers, since the noise spikes then appeared at different wavelengths.

We tested various smoothing techniques on the stellar structure data from model \gasmodel\ as translated to \radmcd . We smoothed three time snapshots simultaneously with equal settings and compared the resulting luminosities, temperatures, and stellar photospheric radius, to obtain an algorithm that is independent of the snapshot. The chosen smoothing schemes were those that, after radiative transfer, resulted in the most similar combination of luminosity, temperature, and radius, as compared to the average values from the \cobold\ simulation.

We limited the smoothing algorithms to work within $1.01\times\,R_*$ ($R_* = 1.65$\,au, Table\,\ref{tab:includedmodels}) as to not introduce any changes to the outer parts of the stellar surface and atmosphere. It affects cells with values that differ by a factor from the median of the surrounding cells (compared from a range of up to ten cells before and after the inspected cell in the \radmcd\ input files). The marked cells are assigned with the surrounding median.

The chosen smoothing algorithm primarily decreased the temperature of a portion of cells well inside the star and produces SEDs with temperatures and luminosities within expected ranges from the simulated star's average values. For example, for the star from \dustmodel , the algorithm decreased the temperature within $< 1.1$\,au, increased the gas density of portion of cells within 0.6 - 1.2\,au, and increased the opacity of a portion of cells within $< 1.3$\,au. This is expected if the problem is due to the Monte Carlo calculations letting photon packages emerge from optical thick and hot regions.

\begin{figure}
    \centering
    \includegraphics[width=90mm]{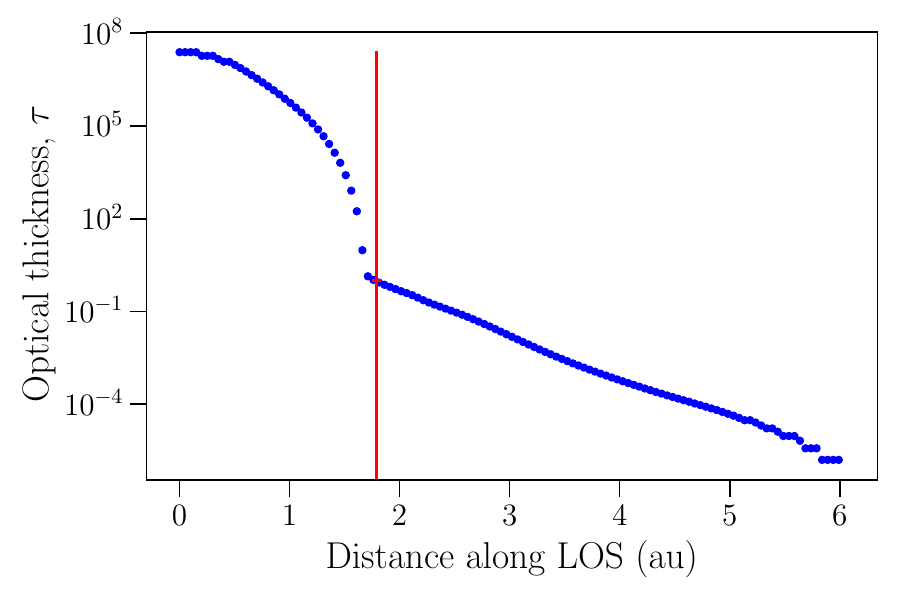}
    \caption{Directional averaged optical thickness of the star and atmosphere of \gasmodel\ as computed with Eq.\,(\ref{eq:optthick}). The red vertical line represents where the optical thickness $\tau = 1$.}
    \label{fig:optthick}
\end{figure}

In Fig.\ref{fig:optthick} we plot the optical thickness through the gas of the smaller model \gasmodel . The optical thickness is given by

\begin{equation}
{\rm d} \tau_{\nu} = - \kappa_{\nu} \rho {\rm d}x = - \chi_{\nu}{\rm d}x
\label{eq:optthick}
\end{equation}
where $\kappa_{\nu}$ is the frequency dependent mass extinction coefficient with unit cm$^2$\,g$^{-1}$, $\rho$ is the mass density in g\,cm$^{-3}$ (both $\kappa_{\nu}$ and $\rho$ are spatially dependent functions of $\vec r$), and d$x$ denotes distance along LOS. The $\chi_{\nu}$ is in turn an extinction coefficient in units cm$^{-1}$. Here, the transition from atmosphere to stellar interior is clearly visible with a sudden increase in optical thickness. With smoothing, $\tau = 1$ falls within $\lesssim 1.1\,R_*$, which is an acceptable discrepancy considering our approximate estimate and the non-spherical nature of the stellar surface.

\subsubsection{Spherically symmetric star from \darwin }

In general, the methods to translate \darwin\ data for usage in \radmcd\ were similar to how we worked with \cobold\ data. The obvious difference is that these data are 1D radial structures.

From the model designated as \darwinmodel , we extracted snapshots of radial gas density, temperature, and opacity in the form of $\kappa_{\rm Ross}$ (again in units of cm$^2$\,$g^{-1}$) that cover one period of the star, or 490 days. From these we chose the time snapshot where the inner radius of the data was most similar to that of our \cobold\ model star. We then adapted this model to obtain a star with similar temperature and luminosity as the star from \dustmodel , as described below.

We computed an effective temperature for the star with Stefan-Boltzmann law to be 2730\,K, based on the stellar radius and luminosity as given in the snapshot's data file's header. In contrast to \cobold , \darwin\ does not simulate the interior of the star and it instead uses a piston to simulate stellar pulsations. As such, we limited the gas temperature data in the interior to not be higher than the star's effective temperature. This way, we obtained a spherically symmetric model star that exhibits an SED with similar luminosity as the star from \dustmodel\ and avoided the inherent problem with leaking photon packages in Monte Carlo radiative transfer as mentioned above in the \cobold\ data description. The remaining parts of the spatial grid used the temperatures from the \darwin\ gas data as is.

Similar to \cobold , we multiplied the gas density with $\kappa_{\rm Ross}$ to obtain the opacity per radial shell. The density-opacity and temperature data are then assigned to the \radmcd\ grid cells according to radial distance from the centre of the grid (grid is described in Sec.\,\ref{sec:griddescript}). We interpolated these data and reordered them in the style of \radmcd 's input files. Exactly as with the \cobold\ gas data, these were included in \radmcd\ as a ``dust species''. This species has zero scattering and $\kappa_{\rm abs} = 1$, since the density in these data include both mass density and $\kappa_{\rm Ross}$ (see above).

\subsection{Dust density and temperature}
\label{sec:dustdescript}

The dust density output from \cobold\ specifies the total number density of all monomers contained in dust grains at a given location. Multiplying these number densities with the monomer mass gives the dust mass density distribution. In our model we had \mgsio , which has the monomer mass of approximately 140.69\,u, or $2.3362\times 10^{-22}$\,g. Similarly as with the gas data, we looped through both grids and saved the average density of \cobold\ cells that fit within each \radmcd\ cell. For the averaging, we ignored \cobold\ cells with no dust content as to not risk introducing soft edges on the inside of the dust-formation zone. However, this slightly increased the total dust mass since this edge was at $\sim 3.4$\,au from the grid centre at this snapshot, and this was just above the radial limit to the final grid refinement in \radmcd\ (3.3\,au).

The dust was assigned the gas temperature from \cobold\ as in the original simulations. This translation to \radmcd\ has been described in Sect.\,\ref{sec:stardescript}. The only difference here is that we set the temperature to zero in cells without dust.

\subsection{Grain sizes and dust opacities}
\label{sec:opacdescript}

For the purpose of computing dust opacities (radiative cross sections, etc.), in general, the sizes of the dust grains need to be known. They can easily be derived from the known total dust monomer number density (see above) using the known number density of dust grains (given by grain abundance, which is a model parameter, and gas number density), and the bulk density of the dust material (defining the volume occupied by a monomer in a grain; see \citealt{hoefner2008aa,hoefner2008psvt} for details). 

To handle the spatial dependence of grain sizes we split the dust density into separate species. From \cobold , we obtained one grain size per spatial grid cell. This overall range of sizes was divided into a number of grain size bins, as discussed below. The bins were introduced as separate dust species in \radmcd\ that each occupy a unique region of space around the star (we note, however, that \radmcd\ does support spatially overlapping dust species). Without binning the grain sizes we would have the same number of grain sizes as the number of dust-filled grid cells. This was not feasible for our radiative transfer since we included wavelength dependent opacity for each grain size. The questions were then; how many grain size bins (species) to use, and how to distribute the sizes.

For the \cobold\ model \dustmodel\ we obtained a grain size range from 1.3\,nm to a maximum of 1\,\um . This size range is in agreement with observations of dust around AGB stars \citep[e.g.][]{norris2012,ohnaka2016,ohnaka2017}. The optimal range of grain sizes for wind-driving is 0.1 to 1\,\um , leading to efficient scattering of light around the flux maximum of an AGB star \citep{hoefner2008aa}.

\begin{figure}
    \centering
    \includegraphics[width=90mm]{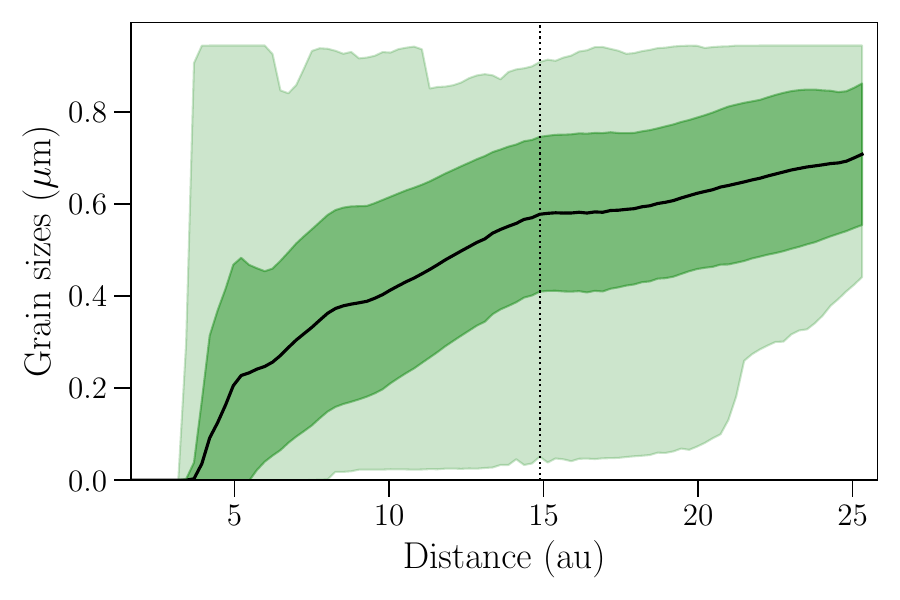}
    \caption{Grain sizes from \dustmodel\ against radial distance from the centre of the \radmcd -grid in 100 spherical shells. The plot starts approximately at the stellar surface, 1.65\,au. The middle black curve is the average grain size of each shell, the darker green field is the standard deviation of the grain sizes, and the lighter and widest green field is the minimum to maximum grain size range. The black vertical dots indicate half the size of the computational box (centre to edge along an axis).}
    \label{fig:grainsizeradius}
\end{figure}

\begin{figure}
    \centering
    \includegraphics[width=90mm]{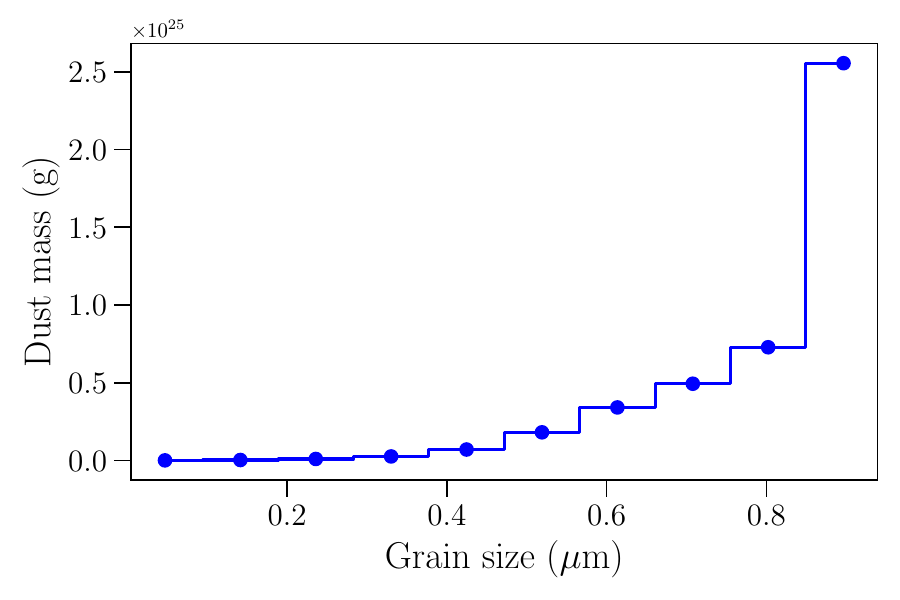}
    \caption{Total mass of each grain size bin for \dustmodel . Symbols mark the exact grain size of each bin.}
    \label{fig:grainsizehist}
\end{figure}

In Fig.\,\ref{fig:grainsizeradius} we show the radial distribution of grain sizes of \dustmodel . We binned these into ten grain size bins and in Fig.\,\ref{fig:grainsizehist} we show the mass of each bin. There is a significant increase in mass for the largest grain size bin. It is normal, however, that the bulk of the dust mass resides in the largest grains, and this model also exhibits rapid grain growth within clouds close to the star that further increases the amount of large grains. This is visible by the rapid increase of the maximum grain size with distance to the star in Fig.\,\ref{fig:grainsizeradius}. The largest grain size bins are the most important contributors to dust emission. Additionally, spectral coverage is also important to consider when choosing grain size bins. For example, we are currently interested in the wavelength ranges from V to MIR. This is approximately between 0.5 to 50\,\um . At 0.5\,\um\ the dust emission is dominated by grains of size $\lambda / (2\,\pi) \approx 0.08$\,\um , well above the smallest grains in our range.

We used wavelength dependent mass absorption and scattering coefficients ($\kappa_{\rm abs}$ and $\kappa_{\rm scat}$ in cm$^2$\,g$^{-1}$), and average scattering angles (parametrised with $g = \left< \cos{\theta} \right>$) in \radmcd . We opted to use opacities based on Mie theory for spherical grains, with refractive indices for amorphous \mgsio\ by \citet{jaeger2003} as extracted from the database of the Astrophysical Laboratory Group of the AIU Jena\footnote{\url{http://www.astro.uni-jena.de/Laboratory/Database/databases.html}}. We used the recently developed \optool\ by \citet{dominik2021}\footnote{\url{https://github.com/cdominik/optool}} to produce the required opacity files for each grain size of this dust species. For Mie opacities, \optool\ uses the routine by \citet{derooij1984}. The \optool\ manual contains details on how it produces the contents of these files, while the \radmcd\ manual contains more details on how the opacities are included for the radiative transfer simulations.

\begin{figure}
    \centering
    \includegraphics[width=90mm]{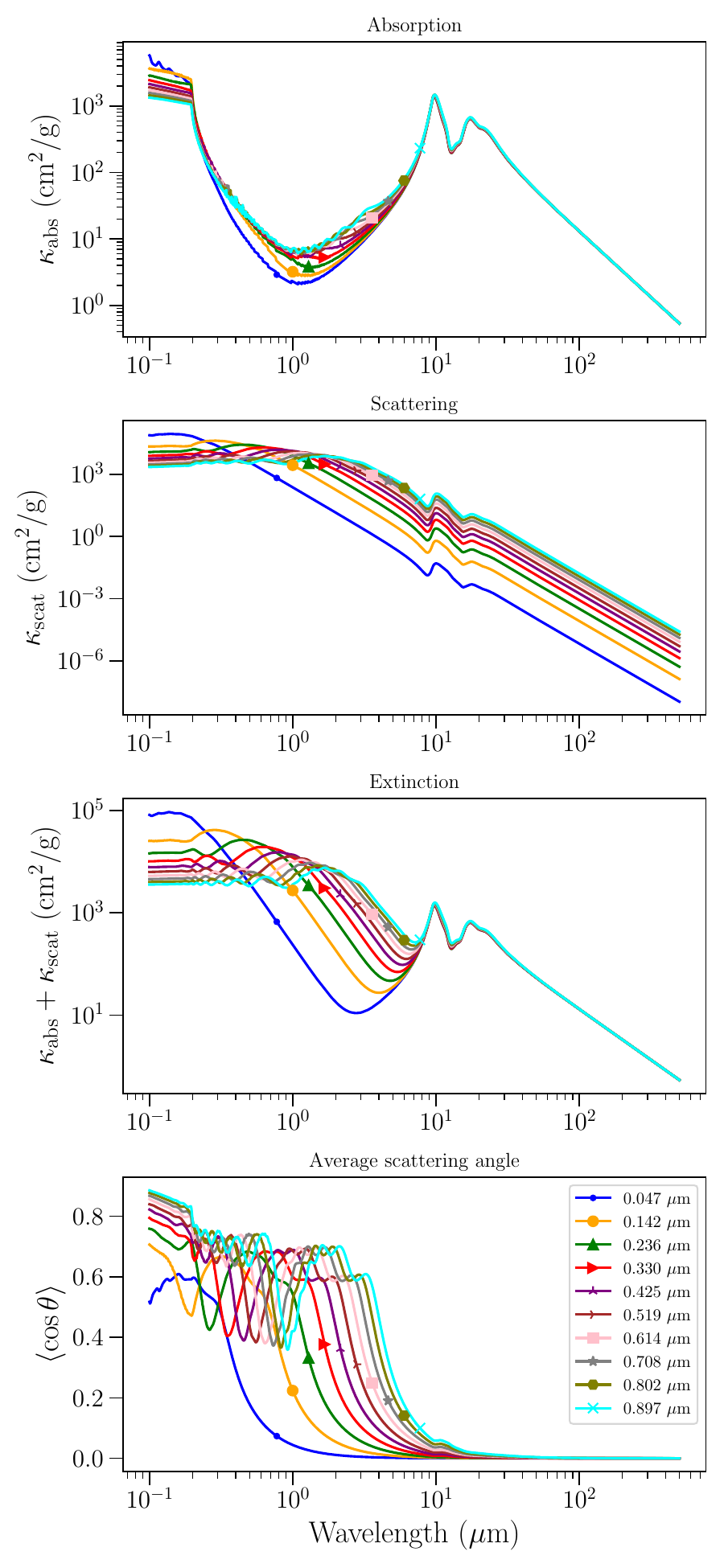}
    \caption{Mass absorptions, scattering, and extinction coefficients and average scattering angles for \mgsio\ dust of each grain size bin.}
    \label{fig:dustopacity}
\end{figure}

We chose to work with ten separate size bins here (i.e. ten species), with linearly separated grain sizes. This is a convenient number of species to use in \radmcd , so that the model is not too demanding to work with. Furthermore, we can see in Fig.\,\ref{fig:dustopacity} that the absorption coefficients of \mgsio , with grain sizes linearly distributed in ten bins between the minimum and maximum are approaching the same values in the MIR regime, just below 10\,\um . This is also where the prominent silicate features appear. A logarithmic grain size binning gives a smaller spread in absorptions, that is, the spectral difference would be small or insignificant. Also, since most of the dust mass is concentrated in the larger grain sizes we considered a linear scale to be suitable. A logarithmic size-scale would concentrate most mass in one grain size bin.

Having fewer than ten grain size bins would be too coarse. For example, with two size bins (smaller than and larger than 0.5\,\um ), the mass ratio would be almost 40, meaning that only 2-3\% of the mass would be in the smaller-sized grains. That is, the spectra would be significantly dominated by the larger grains. The number of grain size bins can of course be changed according to future demands. 

Each bin's grain size was set to be the average of the grain size range that is included in each bin. Considering the small particle limit, we could combine all grains smaller than 0.08\,\um\ into one bin (the largest grains in the smallest bin are then 0.094\,\um ). These final sizes are listed in the legend of Fig.\,\ref{fig:dustopacity}. For the Mie calculations, we divided each grain size bin into 21 grain sizes in a Gaussian distribution centred on the bin's average grain size in a range between the half of the difference between each previous and next grain size bin. The Gaussian shape was such that the full width half maximum (FWHM) coincided with these half-size limits. This softening of grain sizes was done to reduce unrealistic resonances in the opacity that singular-sized spherical grains otherwise create.

\section{Results: Images and spectral energy distributions}
\label{sec:results}

\subsection{Consistent 3D star and dust envelope model}
\label{sec:results:c5dmodel}

\begin{figure*}
    \centering
    \includegraphics[width=160mm]{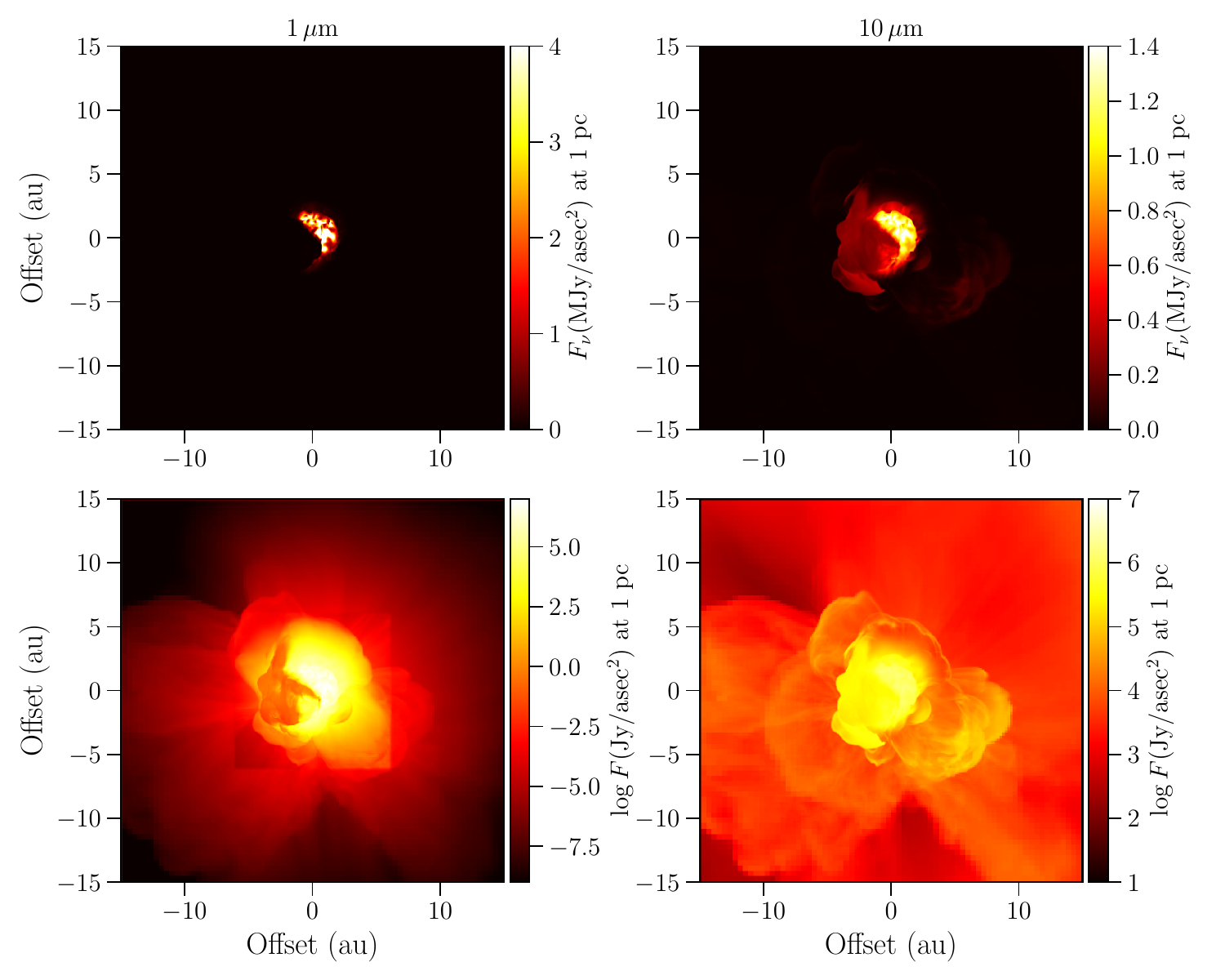}
    \caption{Images created with \radmcd\ from \dustmodel\ at 1 and 10\,\um\ and as seen from the 0-0 angles ($i = 0$\adeg\ and $\phi = 0$\adeg ). The top row has a linear flux scale, and the bottom row has a logarithmic flux scale as indicated by the colour bars where the flux densities are normalised to a distance of 1\,pc.}
    \label{fig:images:examples}
\end{figure*}

Here we show results representative of our SEDs and images as created with \radmcd . In Fig.\,\ref{fig:images:examples} we show images at 1 and 10\,\um\ of \dustmodel\ as seen from the angles $i = 0$\adeg\ and $\phi = 0$\adeg\ (denoted as 0-0 hereafter), featuring a dense dust cloud in the lower left quadrant. This LOS is along the X-axis of the \cobold\ grid (observed from the positive direction). At the wavelength of 10\,\um , one of the strong silicate features appears so the dust emission is easily distinguishable in these images. On the other hand, at the shorter wavelength of 1\,\um , dust is efficient at scattering light, making it appear as dark regions along the LOS in front of the star.

\begin{figure}
    \centering
        \includegraphics[width=90mm]{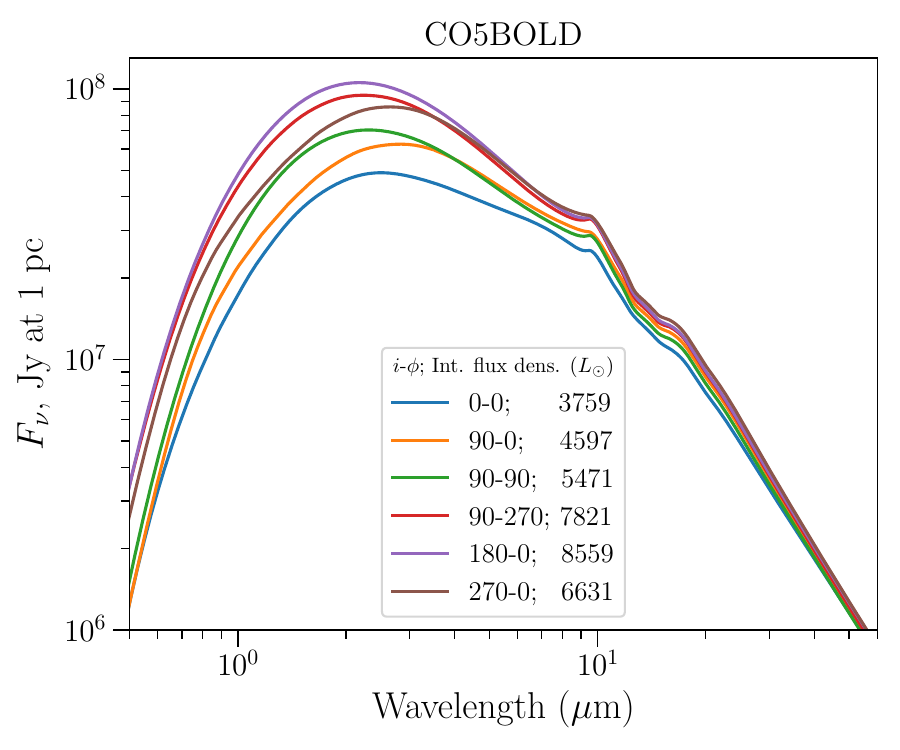}
    \caption{SEDs from \radmcd\ in six viewing directions as indicated by the sets of angles $i$ and $\phi$. Both stellar (gas) and dust data are from \dustmodel . The legend lists the corresponding integrated flux density of each SED.}
    \label{fig:sedc5d}
\end{figure}

In Fig.\,\ref{fig:sedc5d} we show resulting SEDs of \dustmodel\ from \radmcd\ with LOSs from six different directions. In \radmcd , an SED is computed by integrating the total flux density of an image at each wavelength point. That is, these SEDs represent the spatially unresolved case of this model where the ``beam size'' is a $30 \times 30$\,au$^2$ square. The directional average luminosity of these six SEDs is 6140\,\lsol , and the average luminosity of the star itself is 7836\,\lsol .

The SED flux densities vary with observation angle due to changes in both the stellar surface intensity and dust along the LOS. This is most visible in the SED that is seen from the 0-0 observational angle, and in the images shown in Fig.\,\ref{fig:images:examples}. The legend of Fig.\,\ref{fig:sedc5d} lists the integrated flux densities of each direction of the model \dustmodel . With integrated flux density, we mean the total flux density of the whole \radmcd\ beam that is integrated over the total wavelength range in one observed direction. For example, at the 0-0 angle, the integrated flux is the lowest at 3759\,\lsol , while at $i$ and $\phi$ at 180\adeg\ and 0\adeg , the integrated flux is the highest with 8559\,\lsol . Evidently, the 0-0 observational angle is a useful example with plenty of obscuring dust.

For all directions, the long wavelength regime of the SEDs exhibits a spectral index of $\sim 0.5$. This was measured at wavelengths longer than 30\,\um\ to avoid the silicate features.

\subsection{Spherically symmetric star with a 3D dust envelope}
\label{sec:spherecompare}

To demonstrate the importance of using an asymmetric star for radiative transfer of AGB stars, and their surroundings, we exchanged the star from \cobold\ with a spherically symmetric star. This star was based on a 1D model from \darwin\ (\darwinmodel ) with similar size and luminosity as the \cobold\ star, and we kept the surrounding dust density and temperature data from the \cobold\ model \dustmodel . In Appendix\,\ref{app:dusttemperature}, we also compared the \dustmodel\ model with the extreme case of using a (black body) point source as a star within \radmcd\ that was normalised to emit as the other stellar models.

\begin{table}
    \caption{Directional averaged luminosities from \radmcd .}
    \label{tab:luminosities}
    \begin{center}
    \begin{tabular}{cc}
\hline\hline
\noalign{\smallskip}
    Model &  Luminosity (\lsol ) \\
\noalign{\smallskip}
\hline
\noalign{\smallskip}
    \dustmodel\                & 6140  \\
    \darwinmodel\ \&\ dust     & 7737  \\
    Point source \&\ dust      & 7182  \\
\noalign{\smallskip}
    \dustmodel\ without dust   & 7836  \\
    \darwinmodel\ without dust & 7269  \\
\noalign{\smallskip}
\hline
    \end{tabular}
    \end{center}
    \begin{list}{}{}
    \item[Notes.] 
    Average of six directions; $i$ and $\phi$ equals the sets 0-0, 90-0, 90-90, 90-270, 180-0, and 270-0.
    \end{list}
\end{table}

\begin{figure}
    \centering
        \includegraphics[width=90mm]{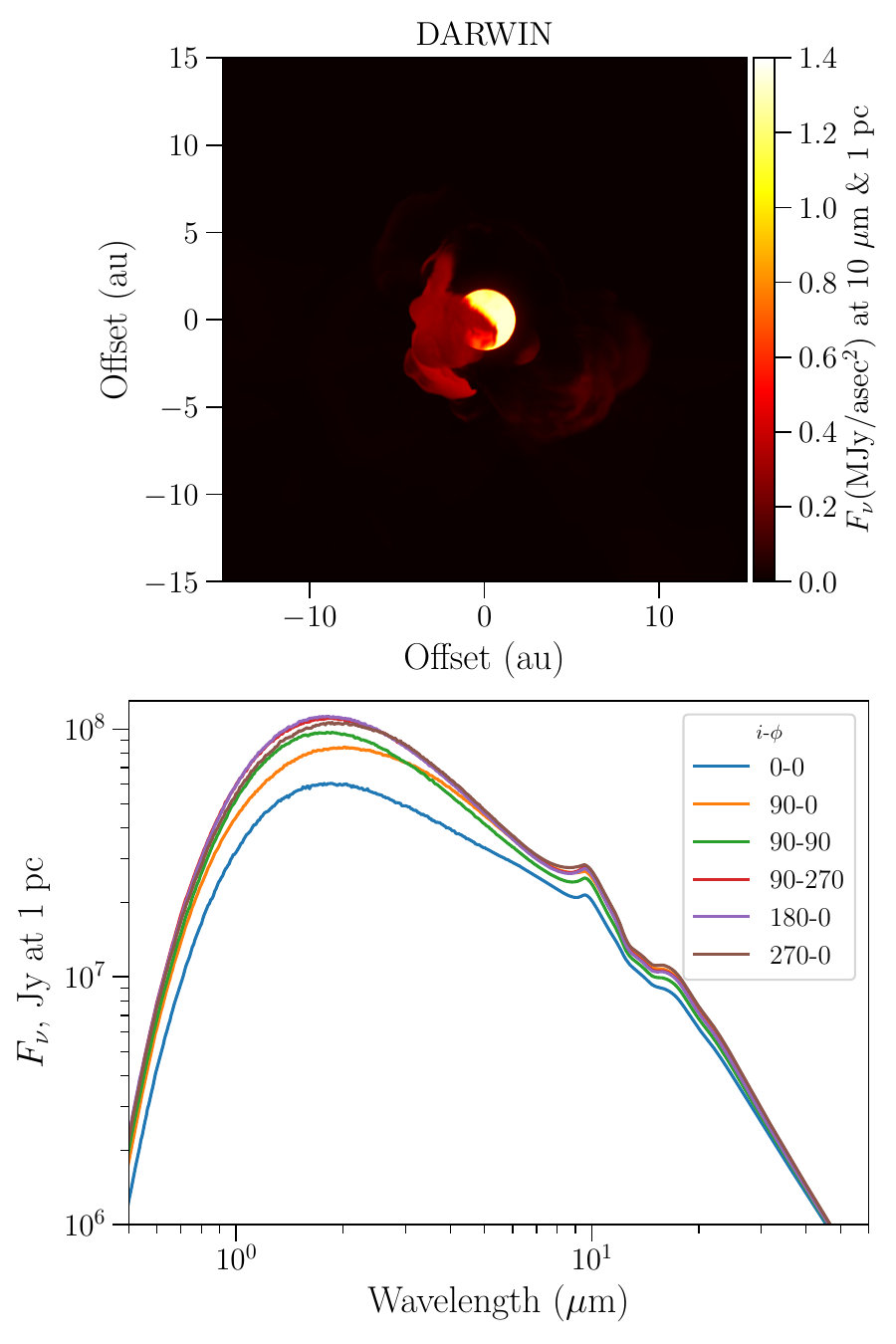}
    \caption{Image and SEDs from \radmcd\ using a star adapted from the model \darwinmodel , and dust density and temperature from the model \dustmodel . The top panel shows a 10\,\um\ image in the 0-0 direction and the bottom panel shows SEDs from \radmcd\ in six viewing directions, as indicated by the sets of angles $i$ and $\phi$.}
    \label{fig:sedimagedarwin}
\end{figure}

In Table\,\ref{tab:luminosities} we present luminosities of these model combinations. For \darwinmodel\ and point source, only the dust distribution varies with observed angle. We see here that the luminosities with isotropic light sources are higher than for \dustmodel . This can be explained with that dust emission and back scattering from dust behind the star is more visible with the star from \darwinmodel , and even more so with a point source. This becomes apparent when we compare the 10\,\um\ images of Figs.\,\ref{fig:images:examples} and Fig.\,\ref{fig:sedimagedarwin}. The surface of the \darwinmodel\ star is much more sharply defined and there is a clear gap between the stellar surface and the first dust clouds. The \dustmodel\ star, on the other hand, is surrounded by more optically thick asymmetric gaseous material, obscuring dust behind the star.

\begin{figure}
    \centering
        \includegraphics[width=90mm]{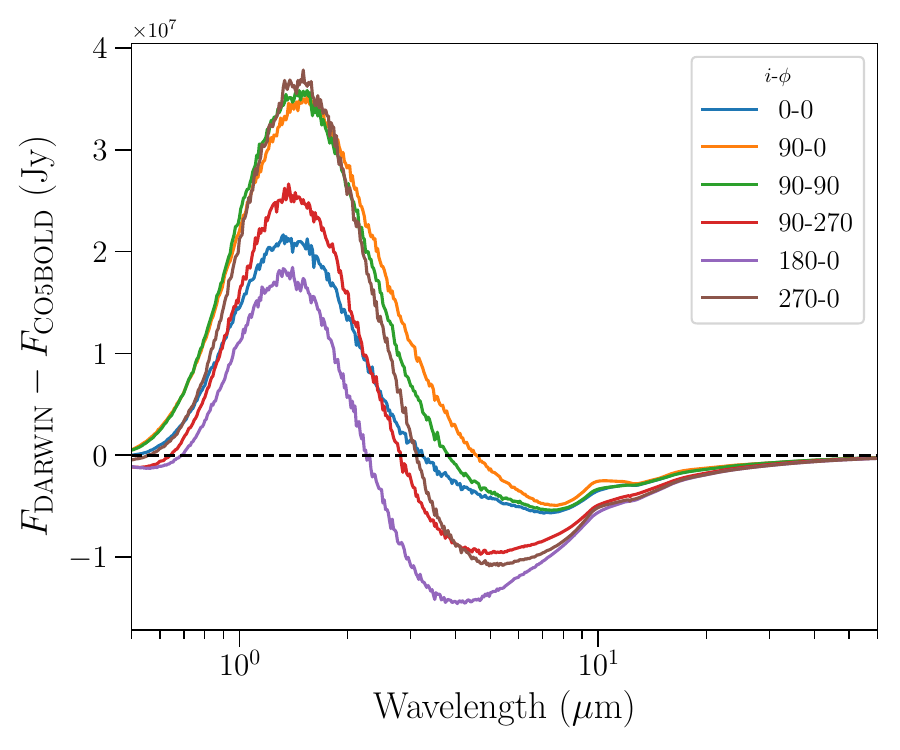}
    \caption{Differences between SEDs simulated with the star from \dustmodel\ ($F_{\rm CO5BOLD}$) and the star from \darwinmodel\ ($F_{\rm DARWIN}$) at a distance of 1\,pc. Both stars are surrounded by dust from \dustmodel . The six viewing directions are indicated by the sets of angles $i$ and $\phi$.}
    \label{fig:sedcompare}
\end{figure}

In Fig.\,\ref{fig:sedimagedarwin} we also show SEDs corresponding to the \darwinmodel\ star surrounded by \dustmodel\ dust, and in Fig.\,\ref{fig:sedcompare} we plot the differences between these SEDs (denoted $F_{\rm DARWIN}$) compared to the \dustmodel\ SEDs (denoted $F_{\rm CO5BOLD}$). Mostly, $F_{\rm DARWIN}$ is higher than $F_{\rm CO5BOLD}$ at wavelengths $\lesssim 2$ - 4\,\um , and the opposite for wavelengths $\gtrsim 2$ - 4\,\um . This is consistent with the interpretation that back scattering on dust close to the star is more visible with the \darwinmodel\ star, since scattering mainly dominates the dust SED at the shorter of these wavelengths. Additionally, we note that these differences are similar in size to the differences in flux densities that depend on observational angles, that is up to a few 10\,MJy at a 1\,pc distance, for both $F_{\rm DARWIN}$ and $F_{\rm CO5BOLD}$. Since $F_{\rm DARWIN}$ uses a spherically symmetric star, this is also an indication that these differences are mainly due to the surrounding dust.

\begin{table}
    \caption{SED peak wavelengths.}
    \label{tab:sedpeak}
    \begin{center}
    \begin{tabular}{ccc}
\hline\hline
\noalign{\smallskip}
    Angles & CO5BOLD & DARWIN \\
     $i$-$\phi$ & \multicolumn{2}{c}{$\lambda_{\rm peak}$ (\um )}  \\
\noalign{\smallskip}
\hline
\noalign{\smallskip}
     0-0   &   2.51  &  1.86  \\
    90-0   &   2.83  &  2.03  \\
    90-90  &   2.30  &  1.86  \\
    90-270 &   2.23  &  1.82  \\
    180-0  &   2.17  &  1.75  \\
    270-0  &   2.64  &  1.86  \\
\noalign{\smallskip}
\hline
\noalign{\smallskip}
  Average  &  2.45  &  1.86  \\
  Max. difference  &  0.66  &  0.27  \\
\noalign{\smallskip}
\hline
    \end{tabular}
    \end{center}
\end{table}

We also note a shift in wavelengths that correspond to the SED's peak flux density. In Table\,\ref{tab:sedpeak} we list each peak wavelength for each observational angle for both \dustmodel\ and \darwinmodel\ plus dust. On average, the $F_{\rm DARWIN}$ peaks are shifted by 0.59\,\um\ to shorter wavelengths. Furthermore, the wavelength of the flux maximum varies much more with viewing angle for the case with the asymmetric star (see Table\,\ref{tab:sedpeak}, and Figs.\,\ref{fig:sedc5d} and \ref{fig:sedimagedarwin}). Since the dust envelope is the same for both model configurations, this can be understood as an effect of angle-dependent illumination, caused by variations of the local stellar surface temperature of the asymmetric star in the 3D model. In summary, the 3D morphology of the dust envelope mainly affects the overall flux levels for the different viewing angles, whereas the illumination by the asymmetric star leads to pronounced wavelength shifts of the flux maximum.

We also compared the average size of the main source at a few wavelengths between 1 and 10\,\um . The main source was defined as the star plus dust emission that would possibly be indistinguishable from the star at a lower angular resolution. The size was set to be the radius where the flux density of an annulus (with $\sim$1 pixel width) is 10\,per\,cent of the source's maximum flux density. At 1-2\,\um , the main sources of both model combinations are comparable in size, but the \dustmodel\ main source grows more with wavelength so that its radius is 29\,per\,cent larger than the \darwinmodel\ star plus dust combined source at 10\,\um .

These differences in source sizes can be expected since the irregular surface of the \dustmodel\ star gives it a generally larger size than the symmetric star. At shorter wavelengths, there are significant amounts of star-obscuring dust so the sizes of the main sources are more similar, while at longer wavelengths we have a combination of the irregular stellar surface plus dust emission, making the \dustmodel 's main source significantly larger than the main source of the \darwinmodel\ star plus dust combination. In summary, we find that switching the \cobold\ stellar model to a similar -- but spherically symmetric star -- gives significant differences in overall SED, and in source shape and size.

Finally, the spectral index at $\lambda > 30$\,\um\ of $F_{\rm DARWIN}$ is $\sim 0.5$, similar as for $F_{\rm CO5BOLD}$. There is an insignificant difference between the spectral indices from \darwinmodel\ and \dustmodel\ (less than 0.1\,per\,cent).

As a further test, we also exchanged the central star of \dustmodel\ with a black body point source while retaining the surrounding dust. For this case, we also re-computed the dust temperature in radiative equilibrium to compare with the temperature assigned to the dust in the original \cobold\ model (see App.\,\ref{app:dusttemperature}). However, since we immediately saw that the lack of spatial extent of the star can lead to a case of extreme and unrealistic obscuration of the star (at, e.g., the 0-0 angle), this rules out the possibility of using this kind of model for any radiative transfer computations producing synthetic observables of models where dust in vicinity of the star is crucial. Even just a minor (optically thick) dust cloud may block an unrealistic amount of stellar emission. Further discussions on effects due to point sources and differences in dust temperatures can be found in Appendix.\,\ref{app:dusttemperature}.

\section{Applications: Synthetic observables}
\label{sec:observables}

One of the main goals of 3D radiative transfer of an asymmetric AGB star, and through its dusty circumstellar environment, is to do case-by-case comparisons with observations. Here we focus on a proof-of-concept test with synthetic images with the star and dust of model \dustmodel . It can be noted that each \cobold\ model can lead to a significant number of unique observables since we can freely chose observation angle and time snapshot. For this study, we do not yet use observational data to compare with. Instead, we extracted various flux densities and contrasts as initial tests of what is resolvable and detectable in the vicinity of the star. This is of immense interest since it has already been shown that modern interferometers are able to resolve the surface and regions around nearby AGB stars \citep[e.g. ][]{khouri2016, paladini2018, rosalesguzman2024, planquart2024}.

It is important to note that our focus was on observable effects from the stellar surface and dust in the immediate surroundings of the star, in particular in the regions where dust and winds are forming. Our computational box had a size of size of $\sim 30^3$\,au$^3$, while normally, AGB stars are surrounded by envelopes with radii on the order of $10^4$\,au. This means that plenty of cold dust along the LOS was missing in our radiative transfer. However, the density and temperature of the envelope, as we move further from the star, is significantly lower than in the vicinity of the star and would not have strong impact on our resulting images and SEDs since we are focussing on the shorter wavelength regimes of V and NIR. In contrast, in images at 100\,\um\ (not included here), no dust emission is visible around the star since dust that emits at these wavelengths will be far outside our computational box. Nevertheless, one should keep in mind that this may affect how strong some of the observed effects we discuss below would be in reality.

\citet{paladini2018} successfully resolved surface features of the AGB star $\pi^1$\,Gruis, which has a distance of 160\,pc. They observed in the H-band with PIONIER\footnote{Precision Integrated Optics Near-infrared Imaging ExpeRiment\newline \url{https://www.eso.org/sci/facilities/paranal/instruments/pionier/overview.html}} \citep{lebouquin2011pionier} at VLTI at wavelengths close to $\sim 1.6$\,\um\ and with baselines of 90\,m, which gave an angular resolution of $\sim 1.8$\,mas. Today, according to recent documentation,\footnote{VLTI: \texttt{VLT-MAN-ESO-15000-4552} (2023-01-30)\\ PIONIER: \texttt{VLT-MAN-ESO-263601\_v102} (2018-07-04)} VLTI can provide baselines of 130.2\,m with the Unit Telescopes, and baselines as long as 201.9\,m will soon be available with the smaller (movable) Auxiliary Telescopes (ATs).

We opted to use such specifications as inspiration for our current test. Normally, the ATs are used for observing evolved stars (C. Paladini, private communication), so we assumed a baseline of 201.9\,m, but disregarded fundamental effects of interferometry due to beam shape and different combinations of baselines. It is important to note that the beam shape and size of an interferometer vary with each observation and depend on a number of factors, for example, elevation of the source and configuration of the array. For simplicity, we assumed circular Gaussian beams and that the angular resolution equal the FWHM of the Gaussian beam. It should also be noted that detailed comparisons with specific VLTI observations require a more complicated post-processing of the synthetic images.

The shortest wavelength VLTI provides with PIONIER is 1.6\,\um . As for longer wavelengths, we chose 3.5 and 10\,\um , which are available with the L and N-bands of MATISSE\footnote{Multi AperTure mid-Infrared SpectroScopic Experiment\newline \url{https://www.eso.org/sci/facilities/paranal/instruments/matisse/inst.html}} \citep{lopez2022matisse}. A simple way to estimate the angular resolution of an optical interferometer is with

\begin{equation}
\textrm{FWHM} = \frac{\lambda}{2\,B}
\label{eq:vltiresolution},
\end{equation}
where $B$ is the maximum baseline of the observation, and the wavelength $\lambda$ is the central wavelength of the spectral channel. Our FWHMs were then 0.8\,mas at $\lambda = 1.6$\,\um , 1.8\,mas at $\lambda = 3.5$\,\um , and 5.1\,mas at $\lambda = 10$\,\um .

We assumed that the minimum requirement for resolving stellar surface features was that the angular area of the star is equal to at least the total angular surface area of four telescope beams. With the beam sizes given above, a star like the one from model \dustmodel\ would then have to be within the distances of 322\,pc and 1985\,pc, when observed at the wavelengths 10 and 1.6\,\um\ respectively.

Within the distance of 322\,pc, one can find several AGB and RGB stars, for example, the aforementioned and already observed $\pi^1$\,Gruis at 160\,pc; R\,Dor, which may be the closest AGB star with its recently revised distance of $44_{-4}^{+5}$\,pc \citep{andriantsaralaza2022}; and $\gamma$\,Cru, which is possibly the closest RGB star \citep{kiss2003,ireland2004} -- with its distance of 27\,pc \citep{leeuwen2007}. For our examples, we chose to put the \dustmodel\ star at 200\,pc.

\begin{table}
    \caption{Total flux density, contrast, and fractional luminosity of stellar and dust components at 200\,pc and $\lambda = 10$\,\um .}
    \label{tab:dustcontrast}
    \begin{center}
    \begin{tabular}{ccccc}
\hline\hline
\noalign{\smallskip}
              &  Flux density  &  Unit \\
\noalign{\smallskip}
\hline
\noalign{\smallskip}
    $F_\nu ({\rm total})^a$ &  592.7  & Jy \\
    $F_\nu ({\rm star})$          &  492.1  & Jy \\
    $F_\nu ({\rm dust})$          &  323.9  & Jy \\
    $F_\nu^{\rm patch}$(total)    &  53.3   & Jy \\
    $F_\nu^{\rm patch}$(average)  &  243    & mJy\,mas$^{-2}$ \\
    $F_\nu^{\rm patch}$(max)      &  356    & mJy\,mas$^{-2}$ \\
\noalign{\smallskip}
\hline
\noalign{\smallskip}
               &  \multicolumn{2}{c}{Contrasts}  &  \\
\noalign{\smallskip}
\hline
\noalign{\smallskip}
    $L_{\rm dust} / L_\star$                   & \multicolumn{2}{c}{0.016} &  \\
    $F_\nu ({\rm dust}) / F_\nu ({\rm star})$  & \multicolumn{2}{c}{0.658} &  \\
    $F_\nu^{\rm patch}({\rm total}) / F_\nu ({\rm star})$  & \multicolumn{2}{c}{0.112} &  \\
\noalign{\smallskip}
\hline
\noalign{\smallskip}
 Patch extent        & Offset coordinate & Unit \\
\noalign{\smallskip}
\hline
\noalign{\smallskip}
    Horizontal range$^b$ &  5 to 20       &  mas \\
\noalign{\smallskip}
    Vertical range$^b$   &  -12.5 to 2.5  & mas \\
\noalign{\smallskip}
    Patch area       &  225    & mas$^2$ \\
\noalign{\smallskip}
\hline
    \end{tabular}
    \end{center}
    \begin{list}{}{}
    \item[Notes.] 
        $^a$ Total flux of images with obscuration, not the sum of stellar and dust fluxes.
        $^b$ Offset angles of patch edges where flux density is measured at a distance of 200\,pc.
    \end{list}
\end{table}

We are interested in testing the detectability of newly formed dust in the vicinity of the star since our model shows thick clouds within some $\lesssim 7$ - 10\,au from the stellar centre. For comparison, there exist several examples of observations of more distant dust and gas clumps. These can be at hundreds of au from AGB stars \citep[e.g.][]{homan2021, coenegrachts2023}, and even more distant \citep[e.g.][]{decin2020, gottlieb2022}, and have unclear origins. \citet{montarges2023} present data with clumps within distances ranging from a few 10\,au to a few 100\,au, with dust formation seemingly occurring in specific regions, and \citet{khouri2020, velillaprieto2023} both show data with asymmetric or clumpy distributions of dust, gas, and wind formation within some 10\,$R_\star$ from the stars. The questions are under what conditions these clouds are formed (original size and distance to the star, temperature, mass density, dust composition), and how they evolve, expand, and finally dissipate into the ISM. The possibilities to answer these questions depend on if it is possible to observe dust clouds as they are forming.

The 0-0 observational angle was the angle of choice for our test since it exhibited a good example of a major dust cloud nearby, partly around, and in front of the star (see Fig.\,\ref{fig:images:examples}). In Table\,\ref{tab:dustcontrast} we list flux densities at 10\,\um\ and a distance of 200\,pc of the whole images. The dust in the 10\,\um\ image emits at around half the strength of the stellar emission, and the fractional luminosity of the dust is on the order of one per\,cent of the stellar luminosity (of 7287\,\lsol ).

To see if the total flux densities are realistic we can compare with, for example, the M-type AGB stars in Table\,1 of \citet{zhao-geisler2012} that listed their flux densities at 12\,\um . These were for the stars R\,Aql, R\,Aqr, R\,Hya, and W\,Hya. These sources' 12\,\um\ flux densities, when re-normalised to 200\,pc, vary between 460 and 2470\,Jy. \citet{khouri2015} have modelled the dust emission of W\,Hya in detail. They find an SED best fit model with Al$_2$O$_3$ dust close to the star (gravitational bound at $2\,R_\star$) and silicates far above the star ($>40$\,au). The 10\,\um\ dust emission shown in their Figs.\,6, 7, and 8, corresponds to a flux density between 150 to 250\,Jy at a re-normalised distance of 200\,pc. Our total flux densities evidently fall in the same order of magnitude as observations of relatively similar stars and their dust emission.

\begin{figure*}
    \hspace{-2mm}
    \includegraphics[width=188mm]{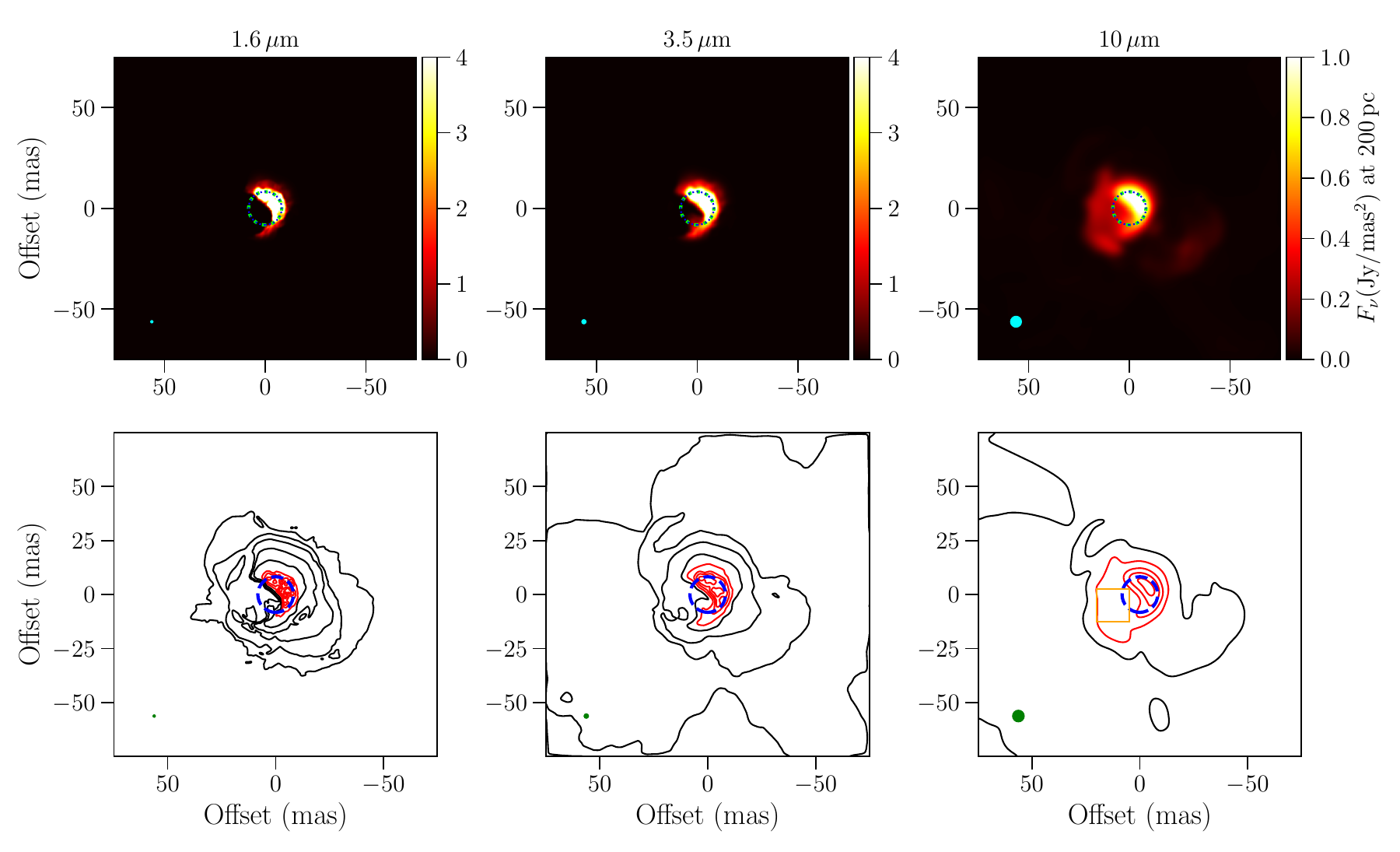}
    \caption{Images and contour plots of the \cobold -model \dustmodel\ simulated with \radmcd\ at a distance of 200\,pc as convolved by Gaussian beams that correspond to the angular resolutions that VLTI can achieve (see text for details). The beam FWHMs are indicated by the cyan (top row) or green (bottom row) circles in the panels' lower left corners. Stellar average radius, 1.65\,au, is indicated with the dotted circle in the centre of the panels. The red contours in the bottom row show flux densities that are $\ge 0.1$ times the image's maximum flux density, and the orange patch in the bottom $\lambda = 10$\,\um\ plot is an area of dust where we measure the patch flux density in Table\,\ref{tab:dustcontrast}.}
    \label{fig:images:vlti}
\end{figure*}

In Fig.\,\ref{fig:images:vlti} we show images and contour plots from \radmcd\ convolved with Gaussians with the FWHMs we calculated above (Eq.\,\ref{eq:vltiresolution}) and at a distance of 200\,pc. In the final contour plot, we show the position of the patch (orange box) from which we extracted flux densities denoted as $F_\nu^{\rm patch}$ in Table\,\ref{tab:dustcontrast}. These numbers are to give an estimate of what flux density level and contrast an observation needs to achieve to detect this cloud.

According to simple tests with the VLTI Exposure time calculator\footnote{\url{https://www.eso.org/observing/etc/bin/gen/form?INS.NAME=MATISSE+INS.MODE=CFP}}, it is possible to reach a signal-to-noise ratio of $\sim 30$ (but 0.45 per frame) when observing a dust disc of 20\,mas diameter, 1000\,K black body, 40\,Jy in N-band (default settings otherwise in the calculator, e.g. 4\,min exposure time). According to the MATISSE documentation, one must also reach a contrast of a factor 0.01 and no less. Our patch's flux density gives a contrast of 0.1 when compared to the total emission of the star at 10\,\um\ (Table\,\ref{tab:dustcontrast}). This specific cloud is clearly within achievable observational limits according to these estimates.

There are plenty of details visible at 1.6\,\um\ of Fig.\,\ref{fig:images:vlti} due to the high angular resolution. The obscuration of the star due to dust is clearly resolvable at the shorter wavelengths, and the dust emission at 10\,\um\ gives the effect of an increased angular size of the central source (as mentioned in Sect.\,\ref{sec:spherecompare}).

These images are not directly comparable to the observations of $\pi^1$\,Gru by \citet{paladini2018}, since they chose this star because of its dust-poor vicinity (so the stellar surface could be more easily distinguished). In its SED (Extended Data Fig.\,3 of \citealt{paladini2018}) there is only dust excess above the stellar SED at wavelengths from 10\,\um\ and longer, and the dust is transparent at shorter wavelengths. Our images are reminiscent of Fig.\,1 of \citet{khouri2016} who observed R\,Dor with the VLT. In their V-band observations the stellar disc is resolved and appears to be significantly obscured. They attribute this to TiO opacity, and we do not include detailed wavelength dependent molecular opacities here. However, the 10\,\um\ panels of our Fig.\,\ref{fig:images:vlti} can be compared with Fig.\,4 of \citet{planquart2024}, who show observations of the C-type AGB star V\,Hya at $\gtrsim 3$ and $\gtrsim 10$\,\um\ (while our model is an M-type star). They see similar asymmetries and blobs around the stellar source at their $\sim\,10$\,\um\ panels, and they attribute these to amorphous carbon (amC) and silicon carbide (SiC) dust emission associated with the star. This is a clearly realistic interpretation of such data.

The remaining question is then whether it is possible to distinguish if such features are due to dust or from stellar surface features (e.g. molecular absorption or a cold patch on the stellar surface caused by convection). One possibility to resolve this question is to observe polarised light due to dust scattering. Another way is by comparing source size at various wavelengths as tested below.

Similarly as in Sect.\,\ref{sec:spherecompare}, we measured the size of the main source by taking average flux densities of circular annuli centred on the middle of the convolved images, and comparing them to the flux density of the brightest pixel of the source. Here, the size of the source was taken to be where the average flux densities are 50, 25, and 10\,per\,cent of the maximum flux density. 

The size ratio between the wavelengths 1.6 and 10\,\um\ is 1.9 and 2.2, with the 25 and 10\,per\,cent limits respectively. The 50\% limit was not useful, however, since the average flux densities of the annuli were too low due to the obscuration of the star. According to \citet{vlemmings2019}, the stellar diameter of an AGB star is only 15-50\% larger at millimetre wavelengths than at the optical. In our test, the average radius of the source is doubled from 1.6 to 10\,\um . Such a significant increase in source size would be difficult to attribute to an increase in photospheric size due to observed wavelengths, rather than the existence a cloud of silicate dust that scatters and emits light close to the star.

\section{Summary and conclusions}
\label{sed:conclusions}

In this paper, we have presented a newly developed method to adapt models obtained with RHD simulations for radiative transfer computations with \radmcd . We used a 3D model produced with \cobold\ of an AGB star and surrounding dust-driven wind within a computation box of the size $\sim 30^3$\,au$^3$. By using this model as a test object, we have produced proof-of-concept observables of the asymmetric star and the dust clouds that appear in the wind-forming regions around the star.

We find that it is possible to simulate an asymmetric evolved star in \radmcd\ by including gas data from \cobold\ simulations as a dust species with zero scattering and an average opacity per grid cell. With \radmcd , we produced SEDs and images with varied observation angles. We used these SEDs and images to extract flux densities and luminosities and to produce synthetic observables.

To demonstrate the importance of using an asymmetric star, we exchanged the star of the 3D model with spherically symmetric stellar models. Primarily, one is based on the star in a 1D model produced with the RHD code \darwin . It turns out that even spatially unresolved observational data should show effects of asymmetries of the star and its surrounding dusty envelope. The synthetic SEDs show a strong dependence on the viewing angle regarding their overall flux levels and shapes. The effect on flux levels is mainly due to the clumpy dust envelope, leading to differences in scattering and extinction of the stellar light in the visual and NIR regime and in thermal emission at mid-IR wavelengths. The variations of the wavelength of the flux maximum with viewing angle, on the other hand, can mostly be attributed to angle-dependent illumination caused by variations of the local stellar surface temperature of the asymmetric star in the 3D model. Clearly, real stars cannot be observed from different viewing angles, but the effects discussed here can be seen as a proxy for time variations at a given viewing angle, as will be explored in more detail in a follow-up paper. The preliminary conclusion, however, is that not all SED variations observed in AGB stars should be interpreted as global changes in the sense of spherical models. Finally, it can be mentioned that using a point source gives extreme and unrealistic obscuration from just small optically thick dust clouds along the LOS and is not suitable for these kinds of models.

We convolved simulated images with artificial (Gaussian) beams to create synthetic images that approximate observations with beam sizes akin to the maximum angular resolutions of the VLTI. From these images, we measured the flux density at 10\,\um\ of a prominent silicate-rich dust cloud to be roughly 200\,mJy\,mas$^{-2}$, at a distance of 200\,pc. The total dust flux density of the whole image at 10\,\um\ (900\,au$^2$, or 22500\,mas$^2$), is roughly 300\,Jy. These numbers correspond well with the observed flux densities we find in the literature. In combination with the flux density contrast to the star at this wavelength, which is $\sim 0.1$, we find that a similar dust cloud should be detectable with VLTI or similar observatories. 

It is also possible to resolve stellar surface features and this dense dust cloud. This is based on the estimated VLTI-based angular resolutions (roughly 200\,m baseline) at the wavelengths of 1.6, 3.5, and 10\,\um\ at 200\,pc.

In our example, the average radius of the central source appears to be approximately twice as large at 10\,\um\ than at 1\,\um\ due to the dense dust cloud along the LOS. This is significantly larger than what one would normally expect when measuring the stellar radius at different wavelengths. Instead, the natural explanation, were this phenomenon to be observed in reality, would be emission from silicate-rich dust in the vicinity of the star.

\begin{acknowledgements}
This work is part of a project that has received funding from the European Research Council (ERC) under the European Union’s Horizon 2020 research and innovation programme (Grant agreement No. 883867, project EXWINGS) and the Swedish Research Council (Vetenskapsrådet, grant number 2019-04059).
The computations were enabled by resources provided by the National Academic Infrastructure for Supercomputing in Sweden (NAISS) and the Swedish National Infrastructure for Computing (SNIC) at UPPMAX partially funded by the Swedish Research Council through grant agreements no. 2022-06725 and no. 2018-05973.
This work has made use of the NumPy library \citep{harris2020}, the SciPy Library \citep{virtanen2020}, the matplotlib library for publication quality graphics \citep{hunter2007}, and the IPython software package \citep{perez2007}.
We thank Claudia Paladini at ESO-VLTI for assistance with information about VLTI.
\end{acknowledgements}

\bibliographystyle{aa}
\bibliography{refs_w24_r3d.bib}

\appendix
\section{Point source and dust temperature}
\label{app:dusttemperature}

\begin{figure}
    \centering
        \includegraphics[width=90mm]{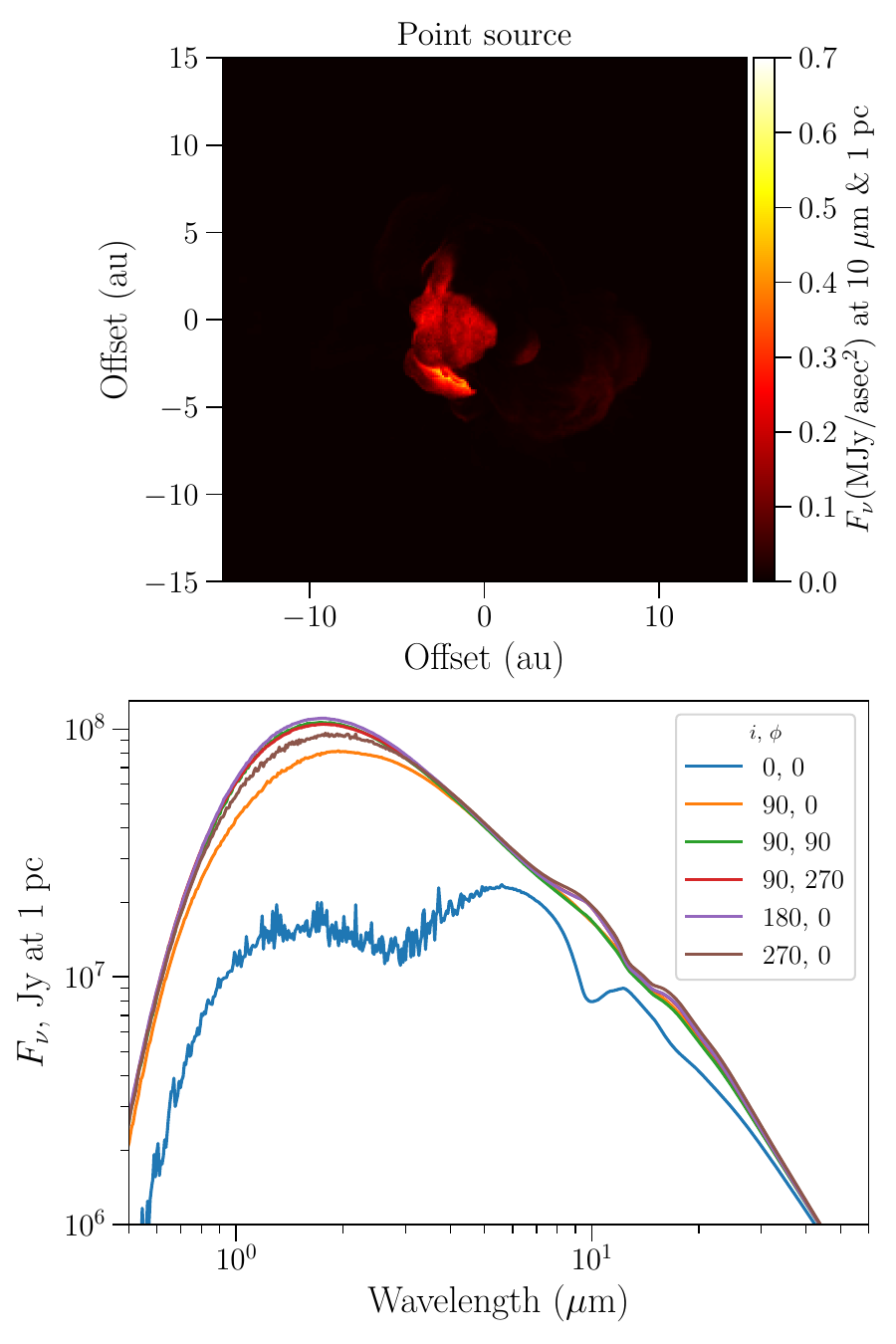}
    \caption{Image and SEDs from \radmcd\ using the dust density from the model \dustmodel , and a central star that is a point source from \radmcd\ that also heats the dust. The top panel shows a 10\,\um\ image in the 0-0 direction. The bottom panel shows SEDs from \radmcd\ in six viewing directions as indicated by the sets of angles $i$ and $\phi$.}
    \label{fig:sedimagepoint}
\end{figure}

In Sect.\ref{sec:spherecompare}, we compared images and SEDs produced by using a spherically symmetric star instead of the asymmetric star from \cobold . Here, we discuss further effects by using a point source within \radmcd\ to re-compute the dust temperature in radiative equilibrium. In Fig.\,\ref{fig:sedimagepoint} we present a 10\,\um\ image of the dust as heated by a point source and the resulting SEDs at various observational angles. Evidently, the 0-0 angle coincides with an extreme case of unrealistic obscuration of the star.

A striking difference between the image in Fig\,\ref{fig:sedimagepoint} and the images of Figs.\,\ref{fig:images:examples} and \ref{fig:sedimagedarwin} is the level of emission by the dust cloud at 10\,\um , despite comparable luminosities of the illuminating stellar sources. The main cause lies in the dust temperatures used in these two cases. In the test scenario, with the spherical star (Fig.\,\ref{fig:sedimagedarwin}), dust is assigned the same temperature as the gas, like in the \cobold\ model. Judging from results of \darwin\ models, where both gas and dust temperatures are computed consistently with non-grey radiative transfer, the temperature of \mgsio\ grains should be lower than the gas temperature (see \citealt{hoefner2008aa}, Fig.2; \citealt{bladh2012}, Fig.6). Being near-transparent at NIR wavelengths, this dust species will not be heated efficiently by the stellar flux, compared to the energy lost through thermal emission in the MIR regime. These non-grey opacity effects are taken into account when computing the dust temperature with \radmcd , leading to a significantly lower dust temperature and less emission at 10\,\um\ in the test case with the point source (Fig.\,\ref{fig:sedimagepoint}). Nevertheless, it has to be kept in mind that a point source with spherically symmetric emission fails to simulate both the effects of the inhomogeneous surface of an AGB star, and the proximity of the stellar surface to the dust-formation zone. Therefore, using this option of \radmcd\ is not a viable solution in this study.

It is important to note here that the translated \cobold\ star cannot be used to compute dust temperatures in \radmcd\ since it requires the input of separate files with data for stellar (symmetric) sources to heat the dust. For this reason, and to stay consistent with the original 3D model, we use the gas temperature computed in \cobold\ for the dust in the main text. This temperature is computed by two approximations. The inner computation box (i.e. the higher resolution box with a side of roughly 12\,au) uses detailed radiation transport including non-grey effects in bins of the frequency range, while the outer computation box uses a power law approximation (see below).

\begin{figure}
    \hspace{-2mm}
        \includegraphics[width=90mm]{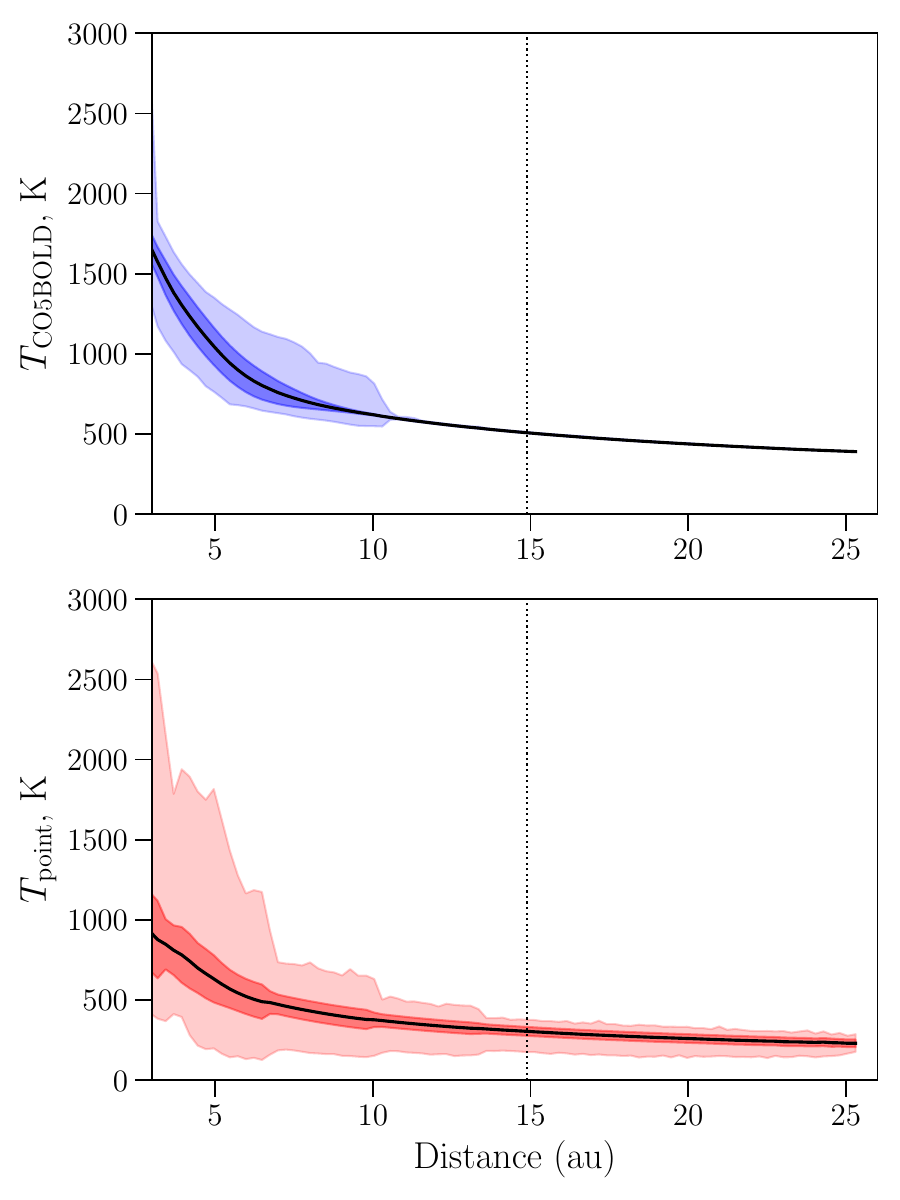}
    \caption{Radial average temperatures of 100 spherical shells of \dustmodel . The top panel shows the temperatures as computed with \cobold . The bottom panel shows the dust temperatures computed in \radmcd\ with a point source at the centre that approximates the star in \dustmodel . The middle line of each curve is the average temperature of each shell, the narrow darker field is the temperature's standard deviation of each shell, and the wide light field show the minimum and maximum temperature of each shell. The plots start at 3\,au, which is just below where dust forms at this snapshot (3.4\,au). The vertical black dots indicate half the edge length (centre to box edge) of the computation box.}
    \label{fig:temperatureradial}
\end{figure}

In Fig.\,\ref{fig:temperatureradial} we compare temperatures as computed with \cobold\ ($T_{\rm CO5BOLD}$) and with \radmcd\ ($T_{\rm point}$) for the dust in \dustmodel . These plots show average temperatures, standard deviations, and minimum-maximum temperature ranges of 100 radial shells. Since there is one temperature per grain size from the point source-computations, the curves in the $T_{\rm point}$ panel show the average, the maximum standard deviation, and minimum-maximum temperature of each grain size of each spherical shell.

The \radmcd\ temperature profile is noisier than the \cobold\ counterpart, since it is based on Monte Carlo simulations, and this propagates to give noisier SEDs than with the \cobold\ temperatures (as can be distinguished in, primarily, the SED of Fig.\,\ref{fig:sedimagepoint}). To significantly decrease the noise we would need increase the number of photon packages by orders of magnitude.

\begin{table}
    \caption{Examples of temperature ratios.}
    \label{tab:temperatureratio}
    \begin{center}
    \begin{tabular}{ccccc}
\hline\hline
\noalign{\smallskip}
      & 3\,au & 6\,au & 9\,au & 15\,au \\
\noalign{\smallskip}
\hline
\noalign{\smallskip}
    $T_{\rm CO5BOLD} / T_{\rm point}$  & 1.81 & 1.65 & 1.63 & 1.67 \\
\noalign{\smallskip}
    $T_{\rm CO5BOLD} / T_{\rm theory}$ & 1.36 & 1.11 & 1.09 & 1.17 \\
\noalign{\smallskip}
    $T_{\rm point}   / T_{\rm theory}$ & 0.75 & 0.67 & 0.67 & 0.70 \\
\noalign{\smallskip}
\hline
    \end{tabular}
    \end{center}
\end{table}

In general, the ratio between the two temperatures is between 1.6 to 1.7, where the \radmcd\ temperature is always lower (see Table\,\ref{tab:temperatureratio}). This is not unexpected since the combination of symmetric point source and non-grey dust opacity both result in lower grain temperatures than in the case of an extended star from the 3D model, with the gas temperature assigned to the dust.

\begin{figure}
    \hspace{-2mm}
        \includegraphics[width=90mm]{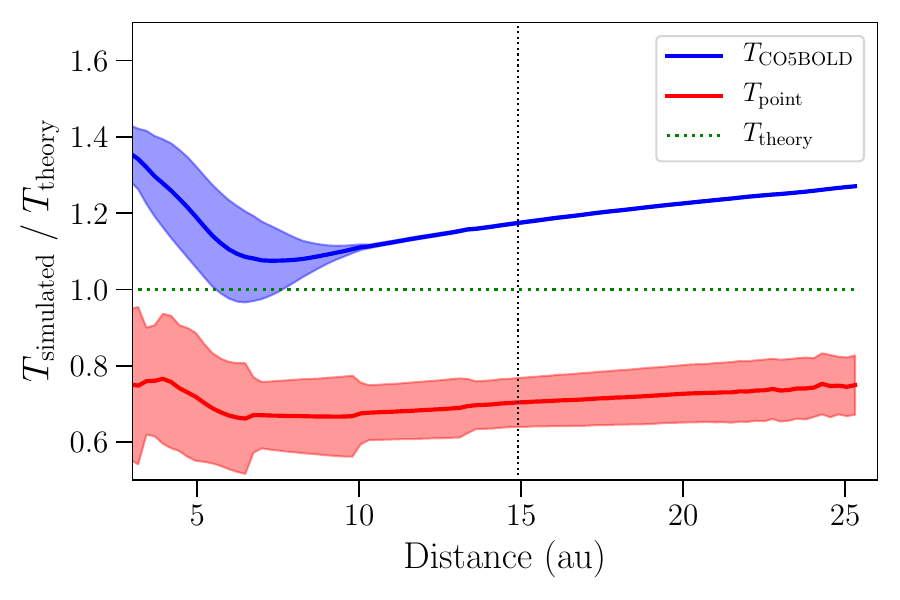}
    \caption{Ratio of dust temperatures used in the radiative transfer to the power-law temperature profile from Eq.\,(\ref{eq:dusttemperature}) (indicated by the green horizontal dotted line). The middle line of each curve is the average value of each shell, and the dark field is the standard deviation of each shell. The plot starts at 3\,au, which is just below where dust forms at this snapshot (3.4\,au). The vertical black dots indicate half the edge length (centre to edge distance) of the computation box.}
    \label{fig:temperatureratio}
\end{figure}

Finally in Fig.\,\ref{fig:temperatureratio}, we also compare these temperatures with a standard formula based on a power-law approximation of the dust opacities. Here, the radial dependence of the dust temperature is formulated as \citep{lamers1999}

\begin{equation}
T_{\rm theory} = T_\star \left( \frac{R_\star}{2\,R} \right)^\frac{2}{4\,+p}
\label{eq:dusttemperature},
\end{equation}
where $p$ is the exponent in a power-law approximation of the wavelength-dependent dust absorption coefficient $\kappa_{\rm abs} \propto\,\lambda^{-p}$ ($p=0$ corresponds to a grey case). For \mgsio\ (using the same refractive index data as in this study), \citet{bladh2012} gave $p = -0.9$ (their Table.\,1), and this we used for the comparison here. This gives a temperature structure that approximates the non-grey effects of the dust, where $T_\star$ is the effective temperature of the star, which is 2800\,K, and $R_\star = 1.65$\,au. For the \dustmodel 's outer computation box (outside a radial distance of $\gtrsim 10$\,au), \cobold\ also assumes this kind of temperature profile but with $p = 0$ \citep[see][for further details]{freytag2023}.

The \cobold\ temperature fits better with the approximate non-grey formula, from the inner edge of the dust condensation zone to $\sim 8$\,au. The \radmcd\ radiative equilibrium temperature appears to approach the simple formula further from the star where the shape, and size, of the star does not contribute significantly to the heating of dust. Generally, \radmcd\ produces a lower dust temperature compared to the analytical approximation within the whole computation box, while \cobold\ is closer to the theoretically estimated temperature in the important regions where dust is forming, and the wind is being initiated, which was the goal of the original 3D RHD model. Further out, where grain growth becomes insignificant and the dynamics is established, the CO5BOLD model predicts a higher dust temperature when it changes to the $p=0$ approximation. 

A comparison can be made here with the discussion on dust temperature by \citet{wiegert2020}. Their dust envelope models span a few 1000\,au, and as such, the star could be approximated as a point source. One of their points, even though their model was not detailed enough to properly include the vicinity of the star, was how dense dust distributions close to the star can significantly affect temperatures far from the star. Here, we show that the stellar size and structure, and the dusty structures close to the star, lead to temperatures that deviate from estimates obtained with the analytical formula in Eq.\,(\ref{eq:dusttemperature}) for a region relatively close to the star. Further work is required to see how this affects temperatures at larger distance.

The higher dust temperature from \cobold\ results in more pronounced silicate features at $\sim 10$ and 20\,\um\ in Fig.\,\ref{fig:sedc5d}. With the lower \radmcd\ temperature, these are fainter, not visible, or even appear as absorption features in the total SED in Fig.\,\ref{fig:sedimagepoint} (similar as the synthetic SEDs by \citealt{wiegert2020}).

While this kind of discussion is interesting in principle, for this study we nevertheless focused on the dust-forming and wind-forming region ($\lesssim 8$\,au). Based on the discussion above, we again concluded that the \cobold\ temperature was more suitable to use in this study since it included effects of the much more detailed stellar surface. However, for future studies we may need to adapt the dust temperature to take non-grey effects into account.

\end{document}